\documentclass[journal,11pt,draftclsnofoot,onecolumn]{IEEEtran}

\usepackage{graphicx}

\usepackage[sort,compress]{cite}

\usepackage{latexsym}
\usepackage{psfrag}
\usepackage{amsthm,amssymb,amscd,amssymb,amsmath,mathrsfs}
\usepackage[noend]{algpseudocode}
\usepackage{multirow}

\theoremstyle{remark}

\newtheorem{thm}{Theorem}
\newtheorem{defn}{Definition}

\newtheorem{prop}{Proposition}

\newtheorem{exmp}{Example}

\newtheorem{rem}{Remark}

\begin{document}

\title{Supervisor Localization of Discrete-Event Systems based on State Tree Structures}

\author{Kai Cai and W.M. Wonham

\thanks{The authors are with the Systems Control Group, Department of Electrical and Computer
Engineering, University of Toronto, 10 King's College Road, Toronto,
Ontario M5S 3G4 Canada. Emails: kai.cai@scg.utoronto.ca, wonham@control.utoronto.ca.}
}

\maketitle

\begin{abstract}
Recently we developed \emph{supervisor localization}, a top-down
approach to distributed control of discrete-event systems in the
Ramadge-Wonham supervisory control framework. Its essence is the
decomposition of monolithic (global) control action into local
control strategies for the individual agents.  In this paper, we
establish a counterpart supervisor localization theory in the
framework of State Tree Structures, known to be efficient for
control design of very large systems. In the new framework, we
introduce the new concepts of local state tracker, local control
function, and state-based local-global control equivalence. As
before, we prove that the collective localized control behavior is
identical to the monolithic optimal (i.e. maximally permissive) and
nonblocking controlled behavior. In addition, we propose a new and
more efficient localization algorithm which exploits BDD
computation. Finally we demonstrate our localization approach on a
model for a complex semiconductor manufacturing system.
\end{abstract}
%


\section{Introduction} \label{Sec1_Intro}

Recently we developed a top-down approach, called \emph{supervisor
localization} \cite{CaiWon_TAC10,CaiWon_IJAMT10}, to the distributed
control of discrete-event systems (DES) in the language-based
Ramadge-Wonham (RW) supervisory control framework
\cite{RamWon:87,SCDES}. We view a plant to be controlled as
comprised of independent asynchronous agents which are coupled
implicitly through control specifications.  To make the agents
`smart' and semi-autonomous, our localization algorithm allocates
\emph{external} supervisory control action to individual agents as
their \emph{internal} control strategies, while preserving the
optimality (maximal permissiveness) and nonblocking properties of
the overall monolithic (global) controlled behavior. Under the
localization scheme, each agent controls only its own events,
although it may very well need to observe events originating in
other (typically neighboring) agents. We call such a scheme
\emph{distributed control architecture}; in a general sense it is
common in the design and implementation of applications like
multi-robot teams and mobile sensor networks (e.g.
\cite{SaFaMu:07}).

Distinct, though related, control architectures are decentralized,
hierarchical, and heterarchical (for recent developments see e.g.
\cite{FenWon:08,SMP:08,SuSchu:10,SuSchu:12}). Both the distributed
and the latter modular approaches aim to achieve efficient
computation and transparent control logic, while realizing
monolithic optimality and nonblocking. With modular supervision,
global control action is typically allocated among specialized
supervisors enforcing individual specifications. By contrast, with
our distributed supervision it is allocated among the individual
active agents (\cite{CaiWon_TAC10,CaiWon_IJAMT10} provide further
discussion of this distinction).

In this paper we continue our investigation of supervisor
localization, but in the (dual) \emph{state-based} framework of DES.
We adopt the recently developed formalism of \emph{State Tree
Structures} (STS) \cite{MaWon:05,MaWon:06TAC}, adapted from
Statecharts \cite{Har:87}, which has been demonstrated to be
computationally efficient for monolithic (i.e. fully centralized)
supervisor synthesis in the case of large systems. Our aim is to
exploit the computational power of STS to solve distributed control
problems in that case as well.

STS efficiently model hierarchical and concurrent organization of
the system state set. The latter is structured as a hierarchical
\emph{state tree}, equipped with modules (\emph{holons}) describing
system dynamics. For symbolic computation, STS are encoded into
predicates. A second feature contributing to computational
efficiency is the use of \emph{binary decision diagrams} (BDD)
\cite{Bry:86}, a data structure which enables a compact
representation of predicates that admits their logical manipulation.
With BDD representation of encoded STS models, the computational
complexity of supervisor synthesis becomes polynomial in the number
of BDD nodes ($|\mbox{nodes}|$), rather than in the `flat' system
state size ($|\mbox{states}|$). In many cases $|\mbox{nodes}| \ll
|\mbox{states}|$, thereby achieving computational efficiency. In
localization, we exploit both these features of STS.

The contributions of this paper are the following. First, we
establish supervisor localization theory in the STS framework:
formulate the distributed control problem, define the notion of
\emph{control cover} for localization, and prove control equivalence
between local controllers and the monolithic one. Compared to
\cite{CaiWon_TAC10}, this state-based localization theory has
several new features: (1) Localization is implemented not by
automata but by a state tracker and control functions (see Section
II); the corresponding notions of \emph{local state tracker} and
\emph{local control function} appear here for the first time; (2)
the new concept of \emph{state-based control equivalence} between
local and monolithic supervision, which differs from the
language-based notion in \cite{CaiWon_TAC10}; (3) an explicit
definition of the \emph{event sets of local controllers}, which
determine inter-agent communication structure. Our second
contribution is a \emph{symbolic localization algorithm} which
computes local controllers via predicates represented by BDDs; this
algorithm is shown to be more efficient than that in
\cite{CaiWon_TAC10}.

Third, the state size reduction brought about by our localization
algorithm can increase the transparency of control logic for large
systems, as compared to the monolithic STS synthesis of
\cite{MaWon:05,MaWon:06TAC}; the latter can produce complex
supervisors with very many BDD nodes. We illustrate this empirical
result by a case study of the industrial example Cluster Tool taken
from \cite{SuSchu:10,SuSchu:12}. Fourth, we extend localization to
the case where component agents may \emph{share events}, thus
achieving greater formal completeness. As seen in
Section~\ref{subsec:share}, a local controller is computed for each
controllable event; when the latter is shared by several agents its
(case-dependent) implementation is spelled out.

We note that there is a different approach, based on ``polynomial
dynamic systems'', to implement the monolithic supervisor by a set
of distributed supervisors with communication \cite{MaGo:09}. The
approach fixes \emph{a priori} subsets of observable events for
individual agents, which may practically rule out the existence
and/or global optimality of the monolithic supervisor.  By contrast,
our localization approach always guarantees existence and global
optimality, and the observation scopes of individual agents will
result automatically as part of the solution. We also note that in
\cite{SePhMaYo:09,SePh:12}, the authors proposed a multi-agent
coordination scheme in the RW framework similar in general terms to
the distributed control architecture of our supervisor localization.
Their synthesis procedure is essentially, however, a combination of
the existing standard RW supervisor synthesis with partial
observation \cite{SCDES} and supervisor reduction \cite{SuWon:04};
and no approach is presented to handle large systems. In this paper
we establish our original supervisor localization in the STS
framework, intended for large complex systems such as Cluster Tool.

The rest of the paper is organized as follows. In
Section~\ref{Sec2_Preli} we provide preliminaries on STS. In
Section~\ref{Sec3_ProFor} we formulate the distributed control
problem. Section~\ref{Sec4_SupLoc} develops the supervisor
localization theory and presents a symbolic localization algorithm
for computing local controllers. In Section~\ref{Sec5_Appli} we
provide the Cluster Tool case study. Finally in
Section~\ref{Sec6_Concl} we state conclusions.


\section{Preliminaries on State Tree Structures} \label{Sec2_Preli}

This section provides relevant preliminaries on the STS-based
supervisory control theory, summarized from
\cite{MaWon:05,MaWon:06TAC}.

A \emph{state tree structure} (STS) $\textbf{G}$ for modeling DES is
a $6$-tuple:
\begin{align} \label{eq:sts_tree}
\textbf{G} = (\textbf{S}T, \mathcal {H}, \Sigma, \Delta,
\textbf{S}T_0, \textbf{S}T_m).
\end{align}
Here $\textbf{S}T$ is the \emph{state tree} organizing the system's
state set into a hierarchy; $\mathcal {H}$ is the set of
\emph{holons} (finite automata) matched to $\textbf{S}T$ that
describes the `local' behavior of $\textbf{G}$; $\Sigma$ is the
finite event set, partitioned into the controllable subset
$\Sigma_c$ and the uncontrollable subset $\Sigma_u$.  Let $\mathcal
{S}T(\textbf{S}T)$ denote the set of all \emph{sub-state-trees} of
$\textbf{S}T$. Then $\Delta : \mathcal {S}T(\textbf{S}T) \times
\Sigma \rightarrow \mathcal {S}T(\textbf{S}T)$ is the `global'
transition function;
$\textbf{S}T_0 \in \mathcal {S}T(\textbf{S}T)$ is the \emph{initial}
state tree; and $\textbf{S}T_m \subseteq \mathcal {S}T(\textbf{S}T)$
is the set of \emph{marker} state trees. A special type of
sub-state-tree of $\textbf{S}T$ is the \emph{basic (state) tree},
each of which corresponds to one `flat' system state in the RW
framework. Let $\mathcal {B}(\textbf{S}T) \subseteq \mathcal
{S}T(\textbf{S}T)$ be the set of all basic trees of $\textbf{S}T$. A
\emph{predicate} $P$ defined on $\mathcal {B}(\textbf{S}T)$ is a
function $P:\mathcal {B}(\textbf{S}T)\rightarrow \{0,1\}$ where $0$
(resp. $1$) stands for logical `false' (resp. `true').  The
predicate \emph{false} (\emph{true}) is identically $0$ ($1$). Thus,
$P$ can be identified by the subset $B_P$ of basic trees $B_P := \{b
\in \mathcal {B}(\textbf{S}T)\ |\ P(b)=1\}$. We shall often write $b
\models P$ for $P(b)=1$. Also for a sub-state-tree $T \in \mathcal
{S}T(\textbf{S}T)$, we define $T \models P$ if and only if $(\forall
b \in \mathcal {B}(T)) b \models P$. Given the \emph{initial
predicate} $P_0$ with $B_{P_0} := \{b \in \mathcal {B}(\textbf{S}T)\
|\ b \models P_0\} = \mathcal {B}(\textbf{S}T_0)$, and the
\emph{marker predicate} $P_m$ with $B_{P_m} := \{b \in \mathcal
{B}(\textbf{S}T)\ |\ b \models P_m\} = \bigcup_{T \in \textbf{S}T_m}
\mathcal {B}(T)$, the STS $\textbf{G}$ in (\ref{eq:sts_tree}) can be
rewritten as
\begin{align} \label{eq:sts_pred}
\textbf{G} = (\textbf{S}T, \mathcal {H}, \Sigma, \Delta, P_0, P_m).
\end{align}

Next write $Pred(\textbf{S}T)$ for the set of all predicates on
$\mathcal {B}(\textbf{S}T)$, and define propositional logic
connectives for its elements as follows: for every $P,P' \in
Pred(\textbf{S}T)$ and $b \in \mathcal {B}(\textbf{S}T)$, (i) $b
\models (\neg P)$ iff $\neg (b \models P)$; (ii) $b \models (P
\wedge P')$ iff $(b \models P) \wedge (b \models P')$; (iii) $b
\models (P \vee P')$ iff $(b \models P) \vee (b \models P')$.
Introduce for $Pred(\textbf{S}T)$ the partial order $\preceq$
defined by $P \preceq P'$ iff $(\neg P) \vee P'$; namely $P \preceq
P'$ holds exactly when $b \models P \Rightarrow b \models P'$ for
every $b \in \mathcal {B}(\textbf{S}T)$. Under the identification of
$Pred(\textbf{S}T)$ with the power set $Pwr(\mathcal
{B}(\textbf{S}T))$ and $\preceq$ with subset containment
$\subseteq$, it is clear that $(Pred(\textbf{S}T),\preceq)$ is a
complete lattice.  The top element is \emph{true}, the bottom
element \emph{false}.

Important elements in $Pred(\textbf{S}T)$ are the reachability and
coreachability predicates. Let $P \in Pred(\textbf{S}T)$. The
reachability predicate $R(\textbf{G},P)$ holds on just those basic
trees that can be reached in $\textbf{G}$, from some $b_0 \models P
\wedge P_0$, via a sequence of state trees all satisfying $P$.
Dually, the coreachability predicate $CR(\textbf{G},P)$ is defined
to hold on those basic trees that can reach some $b_m \models P
\wedge P_m$ in $\textbf{G}$ by a path of state trees all satisfying
$P$. It holds that $R(\textbf{G},P) \preceq P$ and $CR(\textbf{G},P)
\preceq P$. A predicate $P$ is \emph{nonblocking} (with respect to
$\textbf{G}$) if $R(\textbf{G},P) \preceq CR(\textbf{G},P)$, i.e.
every basic tree reachable from some initial state tree can also
reach some marker state tree in $\textbf{G}$.

Another key property of a predicate is controllability (cf.
controllability of a language \cite{SCDES}). For $\sigma \in \Sigma$
define a map $M_\sigma : Pred(\textbf{S}T) \rightarrow
Pred(\textbf{S}T)$ by $b \models M_\sigma(P)$ iff $\Delta(b,\sigma)
\models P$.  Thus $M_\sigma(P)$ identifies the largest subset of
basic trees from which there is a one-step transition $\sigma$ into
$B_P$, or at which $\sigma$ is not defined (i.e.
$\Delta(b,\sigma)=\emptyset$).  A predicate $P$ is called
\emph{weakly controllable} if $(\forall \sigma \in \Sigma_u)\ P
\preceq M_\sigma(P)$.  Thus $P$ is weakly controllable if it is
invariant under the dynamic flow induced by uncontrollable events.
For an arbitrary predicate $P \in Pred(\textbf{S}T)$ bring in the
family $\mathcal {N}\mathcal {C}(P)$ of nonblocking and weakly
controllable subpredicates of $P$, $\mathcal {N}\mathcal {C}(P):=\{K
\preceq P \ |\ K \mbox{ is nonblocking and weakly controllable}\}$.
Then $\mathcal {N}\mathcal {C}(P)$ is nonempty (since $K = false$
belongs) and is closed under arbitrary disjunctions $\vee$; in
particular the supremal element sup$\mathcal {N}\mathcal {C}(P) :=
\bigvee\{K \ |\ K \in \mathcal {N}\mathcal {C}(P)\}$ exists in
$\mathcal {N}\mathcal {C}(P)$.

Now define a \emph{state feedback control} (SFBC) $f$ to be a
function $f: \mathcal {B}(\textbf{S}T) \rightarrow \Pi$, where $\Pi
:=\{\Sigma' \subseteq \Sigma \ |\ \Sigma_u \subseteq \Sigma'\}$.
Thus $f$ assigns to each basic tree $b$ a subset of events that
always contains the uncontrollable events.  For $\sigma \in \Sigma$
define a \emph{control function} (a predicate) $f_{\sigma} :
\mathcal {B}(\textbf{S}T) \rightarrow \{ 0,1 \}$ according to
$f_{\sigma}(b) = 1$ iff $\sigma \in f(b)$.  Thus the control action
of $f$ is fully represented by the set $\{ f_{\sigma} | \sigma \in
\Sigma \}$. By definition $f_\sigma(\cdot) = true$ for every
uncontrollable event $\sigma$. The closed-loop STS formed by
$\textbf{G}$ and $f$ is then written as
\begin{align} \label{eq:sts_closeloop}
\textbf{G}^f = (\textbf{S}T, \mathcal {H}, \Sigma, \Delta^f, P_0^f,
P_m^f),
\end{align}
where $P_0^f = R(\textbf{G}^f, true) \wedge P_0$, $P_m^f =
R(\textbf{G}^f, true) \wedge P_m$, and the transition function
(under $f$) $\Delta^f(b,\sigma)=\Delta(b,\sigma)$ if
$f_{\sigma}(b)=1$ and $\Delta^f(b,\sigma)=\emptyset$ otherwise. A
SFBC $f$ is \emph{nonblocking} if $R(\textbf{G}^f, true) \preceq
CR(\textbf{G}^f, true)$.

\begin{thm}\label{thm:sts_mono} \cite[Theorem~3.2]{MaWon:05}
Let $P \in Pred(\textbf{S}T)$ and $P_0 \wedge$ sup$\mathcal
{N}\mathcal {C}(P) \neq false$. Then there exists a nonblocking SFBC
$f$ such that $R(\textbf{G}^f,true)= R(\textbf{G},$ sup$\mathcal
{N}\mathcal {C}(P))$.
\end{thm}

Theorem~\ref{thm:sts_mono} is the main result for STS on
synthesizing an optimal (in the sense of supremal, or maximally
permissive) and nonblocking supervisor. The SFBC $f$ in
Theorem~\ref{thm:sts_mono} is represented by the control functions
$f_{\sigma}$, $\sigma \in \Sigma$, defined by
\begin{align} \label{eq:SFBC}
f_{\sigma} := M_\sigma(\mbox{sup}\mathcal {N}\mathcal {C}(P)).
\end{align}
Thus for every $b \in \mathcal {B}(\textbf{S}T)$, $f_\sigma(b)=1$ if
and only if $\Delta(b,\sigma) \models \mbox{sup}\mathcal {N}\mathcal
{C}(P)$.



\begin{figure}[!t]
  \centering
  \includegraphics[width=0.45\textwidth]{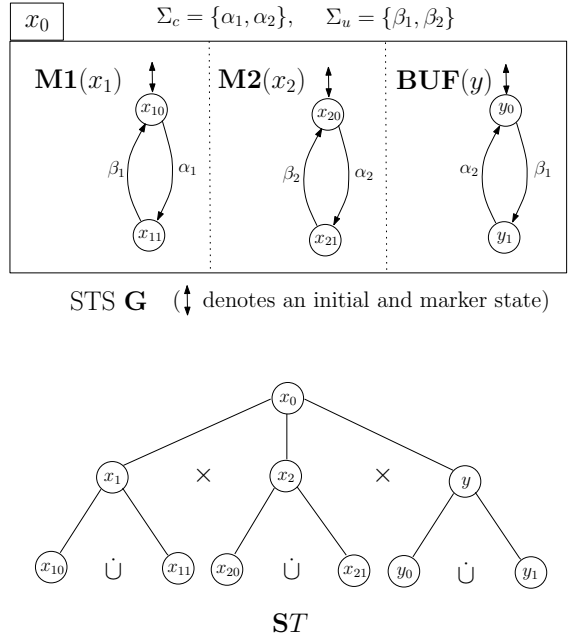}
  \caption{Example: building an STS model. In $\textbf{S}T$, ``$\times$'' denotes cartesian product and ``$\dot{\cup}$'' denotes disjoint union.}
  \label{fig:sts_ex}
\end{figure}

We close this section by describing how to set up a control problem
in STS, as will be needed in Section~\ref{Sec3_ProFor}. Recall
\cite{SCDES} that a finite-state automaton \textbf{P} is defined by
\begin{align} \label{eq:automaton}
\textbf{P} := (Q, \Sigma, \delta, q_0, Q_m),
\end{align}
where $Q$ is the state set, $q_0 \in Q$ is the initial state, $Q_m
\subseteq Q$ is the subset of marker states, $\Sigma$ is the finite
event set, and $\delta: Q \times \Sigma \rightarrow Q$ is the
(partial) state transition function. In the RW (language-based)
framework, a control problem is typically given in terms of a plant
automaton $\textbf{P}$ and a specification automaton $\textbf{S}$
that imposes control requirements on $\textbf{P}$. We can convert
the pair $(\textbf{P},\textbf{S})$ into an STS $\textbf{G}$ with a
predicate $P$ specifying the \emph{illegal} basic trees that
$\textbf{G}$ is prohibited from visiting. Conversion is illustrated
by the example displayed in Fig.~\ref{fig:sts_ex}.  Here the plant
$\textbf{P}$ consists of two `machines' $\textbf{M1}$,
$\textbf{M2}$, and the specification automaton is the buffer
$\textbf{BUF}$ of capacity one. First assign to each of the three
automata a state variable which takes values in the corresponding
state set; then bring in a \emph{root} state $x_0$ which links the
assigned state variables $x_1, x_2, y$ by cartesian product. Thereby
we obtain the STS $\textbf{G}$. Finally we determine the predicate
$P$ for illegal basic trees according to the control requirements
imposed by the specification $\textbf{S}$. In the example
$\textbf{BUF}$ conveys two requirements: (i) disabling event
$\alpha_2$ at state $y_0$ (so the buffer is protected from
underflow) and (ii) disabling $\beta_1$ at $y_1$ (to prevent
overflow). While the disablement of the controllable event
$\alpha_2$ is legal, that of the uncontrollable $\beta_1$ is
illegal. Hence $P = (x_1=x_{11}) \wedge (y=y_1)$, where $\beta_1$ is
defined at $x_{11}$ and $y_1$.




\section{Problem Formulation} \label{Sec3_ProFor}

Consider a plant automaton $\textbf{P}$ (as defined in
(\ref{eq:automaton})) consisting of $n$ component automata
$\textbf{P}_k$, $k=1,\ldots,n$, called `agents'.

\noindent \emph{Assumption 1.} The agents $\textbf{P}_k$,
$k=1,\ldots,n$, are defined over pairwise disjoint alphabets, i.e.
$\Sigma_k \cap \Sigma_j = \emptyset$ for all $k \neq j \in [1,n]$.
For every $k\in[1,n]$ let $\Sigma_k = \Sigma_{c,k} \dot{\cup}
\Sigma_{u,k}$, the disjoint union of the controllable event subset
$\Sigma_{c,k}$ and uncontrollable event subset $\Sigma_{u,k}$. Then
the plant $\textbf{P}$ is defined over $\Sigma := \Sigma_{c}
\dot{\cup} \Sigma_{u}$, where $\Sigma_{c} := \bigcup_{k=1}^n
\Sigma_{c,k}$ and $\Sigma_{u} := \bigcup_{k=1}^n \Sigma_{u,k}$.

Assumption 1 is made in order to simplify the main development and
presentation of results.  In Section~\ref{subsec:share}, below, we
will remove this assumption, and study the case where agents may
share events.

\noindent \emph{Assumption 2.}  A specification automaton
$\textbf{S}$ is defined over $\Sigma$, imposing a behavioral
constraint on $\textbf{P}$.

\begin{figure*}[!t]
  \centering
  \includegraphics[width=0.85\textwidth]{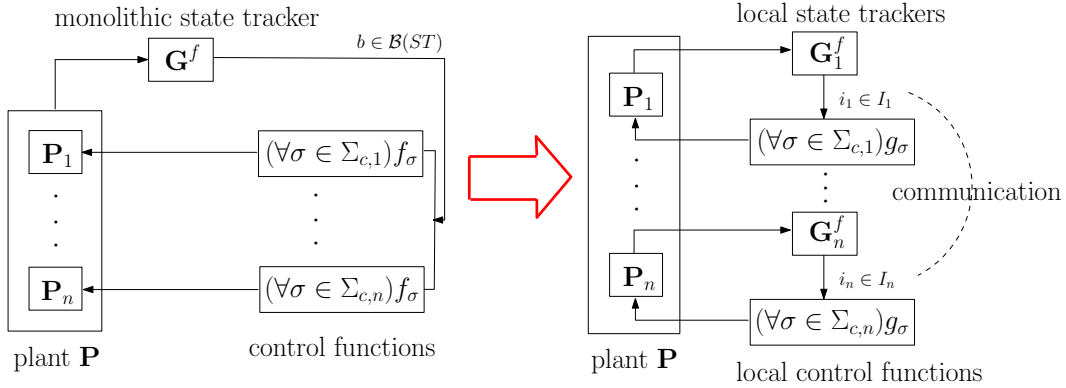}
  \caption{Supervisor localization in STS framework}
  \label{fig:sts_suploc}
\end{figure*}

As stated at the end of Section~\ref{Sec2_Preli}, we convert the
pair ($\textbf{P}, \textbf{S}$) of plant and specification into an
STS $\textbf{G}= (\textbf{S}T, \mathcal {H}, \Sigma, \Delta, P_0,
P_m)$ with a predicate $P$ specifying the illegal basic trees. The
supremal nonblocking and weakly controllable subpredicate of $\neg
P$ is sup$\mathcal {N}\mathcal {C}(\neg P)$, and we suppose
sup$\mathcal {N}\mathcal {C}(\neg P) \wedge P_0 \neq false$ to
exclude the trivial solution. Let
\begin{align} \label{eq:sup}
S := R(\textbf{G}, \mbox{sup}\mathcal {N}\mathcal {C}(\neg P)),\ \ \
B_S := \{ b \in \mathcal {B}(\textbf{S}T) \ | \ b \models S \}.
\end{align}
Then by Theorem~\ref{thm:sts_mono}, there exists a nonblocking SFBC
$f$ (defined in (\ref{eq:SFBC})) such that $R(\textbf{G}^f, true) =
S$, with
\begin{align} \label{eq:P0Pm}
P_0^f = R(\textbf{G}^f, true) \wedge P_0 \ \ \ \mbox{and} \ \ \
P_m^f = R(\textbf{G}^f, true) \wedge P_m.
\end{align}
The SFBC $f$ represented by the control functions $f_{\sigma}$,
$\sigma \in \Sigma$, can be written explicitly as follows:
\begin{align} \label{eq:SFBC2}
(\forall b \in \mathcal {B}(\textbf{S}T))\ f_{\sigma}(b) =\left\{
                                                            \begin{array}{ll}
                                                              1, & \hbox{if either $\Delta(b,\sigma)=\emptyset$ or $\Delta(b,\sigma) \neq \emptyset \ \& \ \Delta(b,\sigma) \models S$;} \\
                                                              0, & \hbox{if $\Delta(b,\sigma) \neq \emptyset \ \& \ \Delta(b,\sigma) \models \neg S$.}
                                                            \end{array}
                                                          \right.
\end{align}
The pair ($\textbf{G}^f, f$) is the monolithic optimal and
nonblocking supervisor for the control problem ($\textbf{G}, P$),
where $\textbf{G}^f$ is the \emph{state tracker} with state set
$B_S$ which supports dynamic evolution of the controlled system, and
$f$ is the SFBC which issues disablement commands based on the state
where $\textbf{G}^f$ currently resides. Since $f$ can be represented
by the set of control functions $\{f_{\sigma}\ |\ \sigma \in
\Sigma_c \}$, the supervisor ($\textbf{G}^f, f$) may be implemented
as displayed on the left of Fig.~\ref{fig:sts_suploc} (cf.
\cite{MaWon:06TAC}). Here the controllable events are grouped with
respect to individual agents $\textbf{P}_k$.

In this implementation, the state tracker $\textbf{G}^f$ is a global
entity, inasmuch as it reports each and every basic tree in $B_S$
that the system visits to all $f_{\sigma}$ for their decision
making. For a purely distributed implementation, we propose to
localize $\textbf{G}^f$ to the individual agents so that each of
them is equipped with its own local state tracker, denoted by
$\textbf{G}^f_k$, $k=1,\ldots,n$.  As will be seen in
Section~\ref{Sec4_SupLoc}, each $\textbf{G}^f_k$ will be constructed
by finding a suitable \emph{cover} $\mathcal {C}_k = \{B_{k,i}
\subseteq B_S \ |\ i \in I_k\}$ on $B_S$; here $B_{k,i}\ (\neq
\emptyset)$ is called a \emph{cell} of $\mathcal {C}_k$, $I_k$ is an
index set, and $\bigcup_{i \in I_k} B_{k,i} = B_S$.  There will also
be a set of marked cells $I_{m,k} \subseteq I_k$. Thus a local state
tracker $\textbf{G}^f_k$ reports system state evolution only in
terms of cells (subsets) of basic trees, rather than singleton basic
trees.
This requires that the associated local control functions
$g_\sigma$, $\sigma \in \Sigma_{c,k}$, take subsets of basic trees
as arguments, i.e. $g_\sigma : Pwr(\mathcal {B}(\textbf{S}T))
\rightarrow \{0,1\}$.  It is then required that $\textbf{G}^f_k$
track exactly the information sufficient for its associated
$g_\sigma$ to issue correct local control. This distributed
implementation is displayed on the right of
Fig.~\ref{fig:sts_suploc}. Finally, we emphasize that in the absence
of monolithic tracking, the local state trackers $\textbf{G}^f_k$
must communicate\footnote{Formally, we consider that communication
is by way of event synchronization; and for simplicity assume that
events are communicated instantaneously, i.e. with no delay.} in
order to give correct reports on system state evolution. The
communication network topology, namely who communicates with whom,
is not given \emph{a priori} but will be generated systematically as
part of our localization result.

\begin{figure}[!t]
  \centering
  \includegraphics[width=0.5\textwidth]{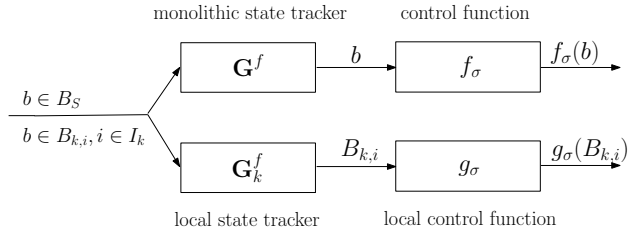}
  \caption{Control equivalence in STS framework}
  \label{fig:con_equiv_sts}
\end{figure}

As usual we require this distributed implementation to preserve the
optimality and nonblocking properties of the monolithic supervisory
control. Fix an arbitrary $k \in [1,n]$ and $\sigma \in
\Sigma_{c,k}$. Suppose that the controlled system is currently
visiting a basic tree $b \in B_S$; then there must exist a cell
$B_{k,i}$, $i \in I_k$, of the cover $\mathcal {C}_k$ to which $b$
belongs. As displayed in Fig.~\ref{fig:con_equiv_sts}, the
monolithic state tracker reports $b$ to $f_\sigma$ which then makes
the control decision $f_\sigma(b)$; on the other hand, a local state
tracker reports the whole cell $B_{k,i}$ to $g_\sigma$ which then
makes the control decision $g_\sigma(B_{k,i})$. We say that the two
pairs $(\textbf{G}^f, f_\sigma)$ and $(\textbf{G}^f_k, g_\sigma)$
are \emph{control equivalent} if for every $b \in B_S$, there exists
$i \in I_k$ such that $b \in B_{k,i}$ (a cell of $\mathcal {C}_k$)
and
\begin{align} \label{eq:con_equiv}
& \Delta(b,\sigma) \neq \emptyset \ \Rightarrow \ \Big[ f_\sigma(b)
= 1
\mbox{ if and only if } g_\sigma(B_{k,i}) = 1 \Big];\\
\label{eq:mark_equiv} & b \models P_m^f \mbox{ if and only if } b
\models P_m \ \& \ i \in I_{m,k}.
\end{align}
Thus (\ref{eq:con_equiv}) requires equivalent enabling/disabling
action, and (\ref{eq:mark_equiv}) requires equivalent marking
action. This form of control equivalence is distinct from the
language-based equivalence in \cite{CaiWon_TAC10}.

We can now formulate the \emph{Distributed Control Problem}. Given a
plant automaton $\textbf{P}$ (as defined in (\ref{eq:automaton})) of
component agents $\textbf{P}_1,\ldots,\textbf{P}_n$ and a
specification automaton $\textbf{S}$ satisfying Assumptions 1 and 2,
let $\mbox{\textbf{SUP}} := (\textbf{G}^f,\{f_\sigma | \sigma \in
\Sigma_c\})$ be the corresponding STS monolithic supervisor, where
$\textbf{G}$ is the STS converted from ($\textbf{P}, \textbf{S}$).
Construct a set of local state trackers $\mbox{\textbf{LOC}}_{st} :=
\{\textbf{G}^f_k | k \in [1,n]\}$, one for each agent, with a
corresponding set of local control functions
$\mbox{\textbf{LOC}}_{cf} := \{g_\sigma, \sigma \in \Sigma_{c,k} | k
\in [1,n]\}$ such that $\mbox{\textbf{LOC}} :=
(\mbox{\textbf{LOC}}_{st},\mbox{\textbf{LOC}}_{cf})$ is
\emph{control equivalent} to $\mbox{\textbf{SUP}}$: that is, for
every $k \in [1,n]$ and every $\sigma \in \Sigma_{c,k}$, the pairs
$(\textbf{G}^f, f_\sigma)$ and $(\textbf{G}^f_k, g_\sigma)$ are
control equivalent in the sense defined in (\ref{eq:con_equiv}) and
(\ref{eq:mark_equiv}).

For the sake of easy implementation and comprehensibility, it would
be desired in practice that the number of cells of local state
trackers be much less than the number of basic trees of their
`parent' monolithic tracker, i.e. $(\forall k \in [1,n])\
|\textbf{G}^f_k| << |\textbf{G}^f|=|B_S|$, where $|\cdot|$ denotes
the size of the argument. Inasmuch as this property is neither
precise to state nor always achievable, it will be omitted from the
formal problem statement; in applications, nevertheless, it should
be kept in mind.


\section{Supervisor Localization} \label{Sec4_SupLoc}

We solve the Distributed Control Problem by developing a supervisor
localization procedure in the STS framework. Although the procedure
is analogous to the development in the RW framework of
\cite{CaiWon_TAC10}, we will formally present the new notions of
local state tracker and local control function, explicitly define
the event sets of local controllers, and provide a new proof which
establishes the state-based control equivalence between local and
monolithic supervision.

\subsection{Construction Procedure} \label{subsec:proc}

We need some notation from \cite{MaWon:05}. Let $\sigma \in \Sigma$
and $P \in Pred(\textbf{S}T)$. Then $\Gamma(P,\sigma)$ is the
predicate which holds on the largest set of basic trees, each of
which can reach a basic tree in $B_P$ by a one-step transition
$\sigma$.  Also $Next_\textbf{G}(\sigma)$ is the predicate which
holds on the largest set of basic trees of $\textbf{G}$ that is
reachable by a one-step transition $\sigma$. Define the legal
subpredicate $N_{good}(\sigma)$ of $Next_\textbf{G}(\sigma)$ by
$N_{good}(\sigma) := Next_\textbf{G}(\sigma) \wedge S$, and the
illegal subpredicate $N_{bad}(\sigma) := Next_\textbf{G}(\sigma)
\wedge \neg S$, where $S$ is the supervisor predicate in
(\ref{eq:sup}).

Now fix an arbitrary $k \in [1,n]$. We develop a localization
procedure which decomposes the monolithic state tracker
$\textbf{G}^f$ into a local state tracker $\textbf{G}^f_k$ for agent
$\textbf{P}_k$ defined over $\Sigma_k$. First, we establish a
\emph{control cover} on $B_S$ (in (\ref{eq:sup})), the state set of
$\textbf{G}^f$, based solely on the control and marking information
pertaining to $\Sigma_{c,k}$, as captured by the following four
functions. Let $\sigma \in \Sigma_{c,k}$. Define $E_{\sigma} : B_S
\rightarrow \{ 0,1 \}$ by
\begin{align} \label{eq:E}
E_\sigma := \Gamma(N_{good}(\sigma), \sigma) \wedge S.
\end{align}
Thus $E_\sigma$ is the characteristic function of the set of basic
trees in $B_S$ where $\sigma$ is enabled. By this definition, for
every $b \in B_S$, $b \models E_\sigma$ if and only if
$\Delta(b,\sigma) \neq \emptyset$ and $f_\sigma(b)=1$ ($f_\sigma$
defined in (\ref{eq:SFBC2})).
Next define $D_\sigma : B_S \rightarrow \{ 0,1 \}$ by
\begin{align} \label{eq:D}
D_\sigma := \Gamma(N_{bad}(\sigma), \sigma) \wedge S.
\end{align}
Namely, $D_\sigma$ is the characteristic function of the set of
basic trees in $B_S$ where $\sigma$ must be disabled by the
supervisory control action of $S$. Thus for every $b \in B_S$, $b
\models D_\sigma$ if and only if $f_\sigma(b)=0$. Also define $M :
B_S \rightarrow \{ 0,1 \}$ according to
\begin{align} \label{eq:M}
M(b)=1 \mbox{ if and only if } b \models P^f_m,\ \ P^f_m \mbox{ in }
(\ref{eq:P0Pm}).
\end{align}
Thus $M$ holds on the set of basic trees which are marked in $B_S$
(i.e. in $\textbf{G}^f$). Finally define $T:B_S \rightarrow \{ 0,1
\}$ according to
\begin{align} \label{eq:T}
T(b)=1 \mbox{ if and only if } b \models P_m,\ \ P_m \mbox{ in }
(\ref{eq:sts_pred}).
\end{align}
So $T$ holds on the set of basic trees originally marked in
$\textbf{G}$. Note that for each $b \in B_S$, we have by $P_m^f =
R(\textbf{G}^f, true) \wedge P_m$ (in (\ref{eq:P0Pm})) that $T(b)=0
\Rightarrow M(b)=0$ and $M(b)=1 \Rightarrow T(b)=1$.  Based on the
above four functions of the control and marking information of
$\Sigma_{c,k}$, we define the following key binary relation
$\mathcal {R}_k$ on $B_S$.
\begin{defn}\label{defn:consis}
Let $\mathcal {R}_k \subseteq B_S \times B_S$. We say that $\mathcal
{R}_k$ is a \emph{control consistency relation} (with respect to
$\Sigma_{c,k}$) if for every $b, b' \in B_S$, $(b, b') \in \mathcal
{R}_k$ if and only if
\begin{align*}
&(i)\ (\forall \sigma \in \Sigma_{c,k})\ E_\sigma(b) \wedge
D_\sigma(b') = false = E_\sigma(b') \wedge D_\sigma(b);\\
&(ii)\ T(b)=T(b') \Rightarrow M(b) = M(b').
\end{align*}
\end{defn}

\begin{figure}[!t]
  \centering
  \includegraphics[width=0.43\textwidth]{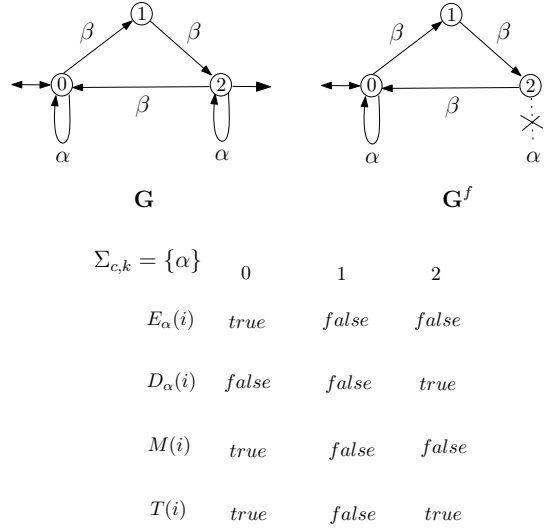}
  \caption{Control consistency relation $\mathcal {R}_k$ is not transitive: $(0,1) \in \mathcal {R}_k$, $(1,2) \in \mathcal {R}_k$, but
  $(0,2) \notin \mathcal {R}_k$.}
  \label{fig:con_consis}
\end{figure}

Informally, a pair of basic trees $(b,b')$ is in $\mathcal {R}_k$ if
there is no event in $\Sigma_{c,k}$ that is enabled at $b$ but is
disabled at $b'$, or vice versa (consistent disablement
information); and (ii) $b$ and $b'$ are both marked or unmarked in
$B_S$ provided that they are both marked or unmarked in $\textbf{G}$
(consistent marking information).  It is easily verified that
$\mathcal {R}_k$ is reflexive and symmetric, but need not be
transitive, and consequently not an equivalence relation (analogous
to \cite{CaiWon_TAC10}); see Fig.~\ref{fig:con_consis}. This fact
leads to the following definition of \emph{control cover}. Recall
that a \emph{cover} on a set $B_S$ is a family of nonempty subsets
(or \emph{cells}) of $B_S$ whose union is $B_S$.
\begin{defn} \label{defn:concov}
Let $I_k$ be some index set, and $\mathcal {C}_k = \{B_{k,i}
\subseteq B_S | i \in I_k\}$ be a cover on $B_S$. We say that
$\mathcal {C}_k$ is a \emph{control cover} (with respect to
$\Sigma_{c,k}$) if
\begin{align*}
&(i)\ (\forall i \in I_k, \forall b,b' \in B_{k,i})\ (b,b') \in \mathcal {R}_k;  \\
&(ii)\ (\forall i \in I_k, \forall \sigma \in \Sigma) \Big[ (\exists
b \in B_{k,i})\ \Delta^f(b,\sigma) \neq \emptyset
\Rightarrow \\
&\hspace{1cm} (\exists j \in I_k)(\forall b' \in B_{k,i})\
\Delta^f(b',\sigma) \subseteq B_{k,j} \Big].
\end{align*}
\end{defn}
A control cover $\mathcal {C}_k$ groups basic trees in $B_S$ into
(possibly overlapping) cells $B_{k,i}$, $i \in I_k$. According to
(i), all basic trees that reside in a cell $B_{k,i}$ have to be
pairwise control consistent; and (ii), for each event $\sigma \in
\Sigma$, all basic trees that can be reached from any basic trees in
$B_{k,i}$ by a one-step transition $\sigma$ have to be covered by a
certain cell $B_{k,j}$ (not necessarily unique). Hence, recursively,
two basic trees $b$, $b'$ belong to a common cell in $\mathcal
{C}_k$ if and only if (1) $b$ and $b'$ are control consistent, and
(2) two future states that can be reached from $b$ and $b'$,
respectively, by the same string are again control consistent. In
the special case where $\mathcal {C}_k$ is a partition on $B_S$, we
call $\mathcal {C}_k$ a \emph{control congruence}.

Having defined a control cover $\mathcal {C}_k$ on $B_S$, we
construct a local state tracker
\begin{align} \label{eq:locstatetrack}
\textbf{G}^f_k = (I_k, \Sigma_{l,k}, \delta_k, i_{0,k}, I_{m,k})
\end{align}
by the following procedure.

\noindent  (P1) Each state $i \in I_k$ of $\textbf{G}^f_k$ is a cell
$B_{k,i}$ of $\mathcal {C}_k$.  In particular, the initial state
$i_{0} \in I_k$ is a cell $B_{k,i_{0}}$ where the basic tree $b_0$
belongs,
i.e. $b_0 \in B_{k,i_{0}}$, and the marker state set $I_{m,k} :=\{i
\in I_k | B_{k,i} \cap \{b \in B_S | b \models P_m^f \} \neq
\emptyset\}$.

\noindent (P2) Choose the local event set $\Sigma_{l,k}$. For this,
define the transition function $\delta'_k: I_k \times \Sigma
\rightarrow I_k$ over the entire event set $\Sigma$ by
\begin{equation} \label{eq:P2}
\begin{split}
\delta'_k(i,\sigma)=j \ \mbox{ if } &(\exists b \in B_{k,i})
\Delta^f(b,\sigma) \neq \emptyset \ \&\ \\
&(\forall b' \in B_{k,i}) \Delta^f(b',\sigma) \subseteq B_{k,j}.
\end{split}
\end{equation}
Choose $\Sigma_{l,k}$ to be the union of $\Sigma_k$ of agent
$\textbf{P}_k$ with events in $\Sigma \setminus \Sigma_k$ which are
\emph{not} selfloop transitions of $\delta'_k$. Thus $\Sigma_{l,k}
:= \Sigma_k \dot{\cup} \Sigma_{com,k}$, where
\begin{align} \label{eq:comm_event}
\Sigma_{com,k} := \{\sigma \in \Sigma \setminus \Sigma_k \ |\
(\exists i,j \in I_k)\ i \neq j \ \& \ \delta'_k(i, \sigma)=j\}.
\end{align}
The set $\Sigma_{com,k}$ determines the subset of agents
$\textbf{P}_j$ ($j \neq k$) that $\textbf{P}_k$ communicates
with.\footnote{The issue of minimal communication in a distributed
system is studied in \cite{RuLaLi:03}.  Although outside the scope
of this paper, minimal communication is an interesting future topic
for localization.}

\noindent (P3) Define the transition function $\delta_k$ to be the
restriction of $\delta$ to $\Sigma_{l,k}$, namely
$\delta_k:=\delta|_{\Sigma_{l,k}} : I_k \times \Sigma_{l,k}
\rightarrow I_k$.

\noindent Thus the above constructed local state tracker
$\textbf{G}^f_k$ is an automaton, which reports system state
evolution in terms of cells (subsets) of basic trees which are
crucial for, and only for, the local control and marking with
respect to $\Sigma_{c,k}$ of agent $\textbf{P}_k$. Owing to the
possible overlapping of cells in $\mathcal {C}_k$, the choices of
$i_0$ and $\delta_k$ may not be unique, and consequently
$\textbf{G}^f_k$ may not be unique. In that case we take an
arbitrary instance of $\textbf{G}^f_k$. Clearly if $\mathcal {C}_k$
happens to be a control congruence, then $\textbf{G}^f_k$ is unique.

Finally, we define \emph{local control functions} $g_\sigma$,
$\sigma \in \Sigma_{c,k}$, to be compatible with $\textbf{G}^f_k$.
Let $\sigma \in \Sigma_{c,k}$. Define $g_\sigma : I_k \rightarrow
\{0,1\}$ by
\begin{align} \label{eq:loc_decmak}
g_\sigma(i)=1 \mbox{\ \ if and only if \ \ } (\exists b \in
B_{k,i})\ b \models \Gamma(N_{good}(\sigma), \sigma).
\end{align}
So $g_\sigma$ will enable $\sigma$ at a state $i$ of the tracker
$\textbf{G}^f_k$ whenever there is a basic tree in the cell
$B_{k,i}$ at which $\sigma$ is enabled.

\begin{figure*}[!t]
  \centering
  \includegraphics[width=0.65\textwidth]{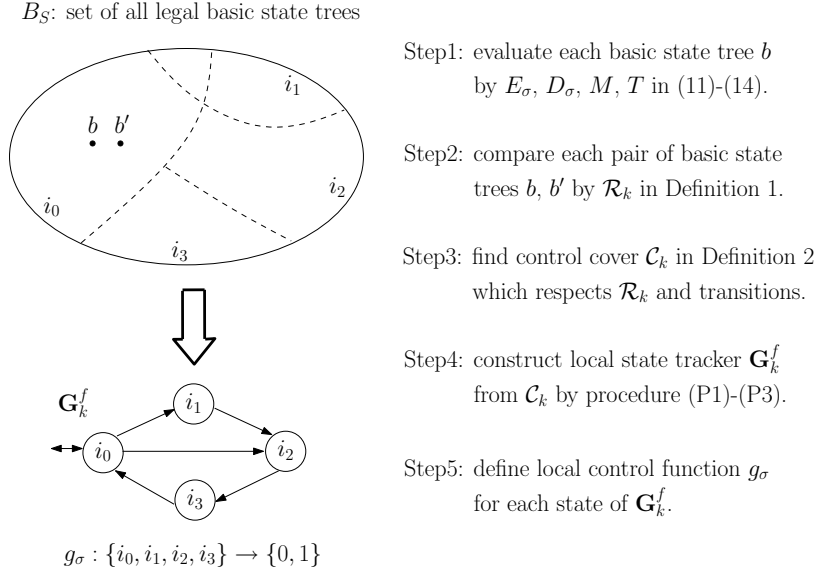}
  \caption{Supervisor localization procedure.}
  \label{fig:Suploc_approach}
\end{figure*}

We have now completed the localization procedure for an arbitrarily
chosen agent $\textbf{P}_k$, $k \in [1,n]$. The procedure is
summarized and illustrated in Fig.~\ref{fig:Suploc_approach}.
Applying the same procedure for every agent, we obtain a set of
local state trackers $\mbox{\textbf{LOC}}_{st} := \{\textbf{G}^f_k |
k \in [1,n]\}$ with a corresponding set of local control functions
$\mbox{\textbf{LOC}}_{cf} := \{g_\sigma, \sigma \in \Sigma_{c,k} | k
\in [1,n]\}$. Our main result, below, states that this pair
$\mbox{\textbf{LOC}} :=
(\mbox{\textbf{LOC}}_{st},\mbox{\textbf{LOC}}_{cf})$ is a solution
to the Distributed Control Problem.

\begin{thm} \label{thm:suploc}
The pair $\mbox{\textbf{LOC}} :=
(\mbox{\textbf{LOC}}_{st},\mbox{\textbf{LOC}}_{cf})$ of local state
trackers and local control functions is control equivalent to the
optimal and nonblocking supervisor $\mbox{\textbf{SUP}} :=
(\textbf{G}^f,\{f_\sigma | \sigma \in \Sigma_c\})$; namely, for
every $k \in [1,n]$, $\sigma \in \Sigma_{c,k}$, and $b \in B_S$,
there exists $i \in I_k$ such that $b \in B_{k,i}$ and
\begin{align*}
& (i)\ \Delta(b,\sigma) \neq \emptyset \ \Rightarrow \ \Big[
f_\sigma(b) = 1
\mbox{ if and only if } g_\sigma(i) = 1 \Big];\\
& (ii)\ b \models P_m^f \mbox{ if and only if } b \models P_m \ \& \
i \in I_{m,k}.
\end{align*}
\end{thm}

The proof, below, establishes the state-based control equivalence
between local and monolithic supervision. It is distinct from, and
more concise than, the language-based proof in \cite{CaiWon_TAC10}.

\noindent \emph{Proof.} Let $k \in [1,n]$, $\sigma \in
\Sigma_{c,k}$, and $b \in B_S$. Then there must exist a state $i \in
I_k$ of the tracker $\textbf{G}^f_k$, corresponding to a cell
$B_{k,i}$ of the control cover $\mathcal {C}_k$, such that $b \in
B_{k,i}$. For (i), suppose that $\Delta(b,\sigma) \neq \emptyset$;
it will be shown that $f_\sigma(b) = 1$ if and only if $g_\sigma(i)
= 1$. (If) Let $g_\sigma(i) = 1$, i.e. there is $b' \in B_{k,i}$
such that $b' \models \Gamma(N_{good}(\sigma), \sigma)$. Since $b'$
is also in $B_S$, we have $b' \models \Gamma(N_{good}(\sigma),
\sigma) \wedge S = E_\sigma$. It follows from $b \in B_{k,i}$ that
$(b,b') \in \mathcal {R}_k$ and $E_\sigma(b') \wedge D_\sigma(b)
\equiv false$. Hence $D_\sigma(b)\equiv false$, which is equivalent
to $f_\sigma(b)=1$ by the definition of $D_\sigma$ in (\ref{eq:D}).
(Only if) Let $f_\sigma(b) = 1$. Since $\Delta(b,\sigma) \neq
\emptyset$ and $b$ is in $B_S$, we have by the definition of
$E_\sigma$ in (\ref{eq:E}) that $b \models E_\sigma =
\Gamma(N_{good}(\sigma), \sigma) \wedge S$. We then conclude from $b
\in B_{k,i}$ and the definition of $g_\sigma$ in
(\ref{eq:loc_decmak}) that $g_\sigma(i) = 1$.

Now we show (ii). (If) Let $b \models P_m$ (i.e. $T(b)=1$) and $i
\in I_{m,k}$. Then there is $b' \in B_{k,i}$ such that $b' \models
P_m^f$; so $M(b')=1$, and also $T(b')=1$.  Since $(b,b') \in
\mathcal {R}_k$ and $T(b)=T(b')$, we have $M(b)=M(b')=1$. (Only if)
Let $b \models P_m^f$ (i.e. $M(b)=1$).  Then $T(b)=1$, i.e. $b
\models P_m$, and also $i \in I_{m,k}$ by the construction of the
tracker $\textbf{G}^f_k$. \hfill $\square$

In essence Theorem~\ref{thm:suploc} asserts that every set of
control covers generates a solution to the Distributed Control
Problem. This raises the converse question: is every solution to the
Distributed Control Problem generated by a suitable set of control
covers? We answer this question in the next subsection.

\subsection{Necessary Structure}

%

Let $\mbox{\textbf{SUP}} = (\textbf{G}^f,\{f_\sigma | \sigma \in
\Sigma_c\})$, with $\textbf{G}^f$ in (\ref{eq:sts_closeloop}) and
$f_\sigma$ in (\ref{eq:SFBC2}), be the monolithic optimal and
nonblocking supervisor for a given control problem.  Also let
$\mathcal {C}_k = \{B_{k,i} | i \in I_k\}$, $k \in [1,n]$ and $I_k$
some index set, be an arbitrary cover on the state set $B_S$ (as in
(\ref{eq:sup})) of $\textbf{G}^f$; namely $\emptyset \neq B_{k,i}
\subseteq B_S$, and $\bigcup_{i \in I_k} B_{k,i} = B_S$. For the
cover $\mathcal {C}_k$ on $B_S$, apply the procedure (P1)-(P3),
above, to obtain an automaton $\textbf{G}^f_k = (I_k, \Sigma_{l,k},
\delta_k, i_{0,k}, I_{m,k})$ as in (\ref{eq:locstatetrack}). We
impose a normality requirement on $\textbf{G}^f_k$ with respect to
$\textbf{G}^f$ (cf. \cite{SuWon:04}).

\begin{defn} \label{defn:norm}
We say that $\textbf{G}^f_k = (I_k, \Sigma_{l,k}, \delta_k, i_{0,k},
I_{m,k})$ with $\delta'_k: I_k \times \Sigma \rightarrow I_k$ in
(\ref{eq:P2}) is \emph{normal} with respect to $\textbf{G}^f =
(\textbf{S}T, \mathcal {H}, \Sigma, \Delta^f, P_0^f, P_m^f)$ if
\begin{align*}
(\forall i \in I_k, \forall \sigma \in \Sigma) &(\exists b \in
B_{k,i})\ \Delta^f(b,\sigma) \neq \emptyset \Rightarrow \\
&(\exists j \in I_k)\ \delta'_k(i,\sigma)=j.
\end{align*}
\end{defn}

\begin{figure}[!t]
  \centering
  \includegraphics[width=0.48\textwidth]{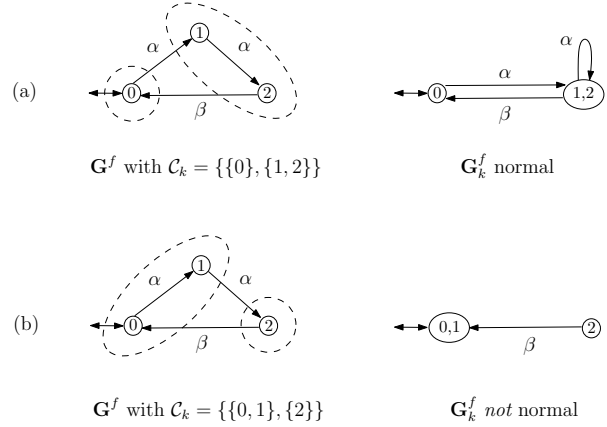}
  \caption{Normality requirement on $\textbf{G}^f_k$ with respect to $\textbf{G}^f$. Here $\textbf{G}^f_k$
  are constructed from cover $\mathcal {C}_k$ on the state set of $\textbf{G}^f$ by the procedure (P1)-(P3).
  In (b) the transitions $\alpha$ are not defined in $\textbf{G}^f_k$ because the condition in (\ref{eq:P2}) is violated for $\sigma=\alpha$.}
  \label{fig:norm}
\end{figure}

Thus normality of $\textbf{G}^f_k$ requires that if an event
$\sigma$ is defined at $b$ $(\in B_{k,i})$ in $\textbf{G}^f$, then
$\sigma$ must be defined by $\delta'_k$ at $i$ $(\in I_k)$ of
$\textbf{G}^f_k$. This requirement in turn imposes a condition on
the cover $\mathcal {C}_k$ from which $\textbf{G}^f_k$ is
constructed, as illustrated in Fig.~\ref{fig:norm}. We will see
below that the condition imposed on $\mathcal {C}_k$ is indeed one
requirement of a control cover.

Now let $\mbox{\textbf{LOC}}_{st} = \{\textbf{G}^f_k | k \in
[1,n]\}$, and $\mbox{\textbf{LOC}}_{cf} = \{g_\sigma, \sigma \in
\Sigma_{c,k} | k \in [1,n]\}$ with $g_\sigma$ defined in
(\ref{eq:loc_decmak}). We say that the pair $\mbox{\textbf{LOC}} =
(\mbox{\textbf{LOC}}_{st},\mbox{\textbf{LOC}}_{cf})$ is
\emph{normal} if every $\textbf{G}^f_k$, $k \in [1,n]$, is normal
with respect to $\textbf{G}^f$. The following result asserts that if
the normal pair $\mbox{\textbf{LOC}} =
(\mbox{\textbf{LOC}}_{st},\mbox{\textbf{LOC}}_{cf})$ is a solution
to the Distributed Control Problem, then the covers $\mathcal
{C}_k$, $k \in [1,n]$, must all be control covers.

\begin{thm}\label{them:sts_1} If the normal pair $\mbox{\textbf{LOC}} =
(\mbox{\textbf{LOC}}_{st},\mbox{\textbf{LOC}}_{cf})$ is control
equivalent to $\mbox{\textbf{SUP}}$, then the covers $\mathcal
{C}_k$, $k \in [1,n]$, are control covers.
\end{thm}
\emph{Proof.} Fix an arbitrary $k \in [1,n]$. According to
Definition~\ref{defn:concov}, we must prove the following two
conditions for cover $\mathcal {C}_k$:
\begin{align*}
&(i)\ (\forall i \in I_k, \forall b,b' \in B_{k,i})\ (b,b') \in \mathcal {R}_k;  \\
&(ii)\ (\forall i \in I_k, \forall \sigma \in \Sigma) \Big[ (\exists
b \in B_{k,i})\ \Delta^f(b,\sigma) \neq \emptyset \Rightarrow \\
&\hspace{1cm} (\exists j \in I_k)(\forall b' \in B_{k,i})\
\Delta^f(b',\sigma) \subseteq B_{k,j} \Big].
\end{align*}

For (ii), let $i \in I_k$, $\sigma \in \Sigma$, and suppose there
exists $b \in B_{k,i}$ such that $\Delta^f(b,\sigma) \neq
\emptyset$. Since $\textbf{G}_k^f$ is normal with respect to
$\textbf{G}^f$, by Definition~\ref{defn:norm} there exists $j \in
I_k$ such that $\delta'_k(i,\sigma)=j$. It then follows from
(\ref{eq:P2}) that $(\forall b' \in B_{k,i})\ \Delta^f(b',\sigma)
\subseteq B_{k,j}$.

Next for (i), let $i \in I_k$ and $b, b' \in B_{k,i}$; it will be
shown that $(b,b') \in \mathcal {R}_k$
(Definition~\ref{defn:consis}).  First, let $\sigma \in
\Sigma_{c,k}$; if $\Delta(b,\sigma)=\emptyset$ (resp.
$\Delta(b',\sigma)=\emptyset$), then $E_\sigma(b)=D_\sigma(b)=false$
by (\ref{eq:E}) and (\ref{eq:D}) (resp.
$E_\sigma(b')=D_\sigma(b')=false$). Hence there holds $E_\sigma(b)
\wedge D_\sigma(b')=false=E_\sigma(b') \wedge D_\sigma(b)=false$ if
$\sigma$ is not defined at $b$ or $b'$ or at both of them. Now
suppose that $\Delta(b,\sigma)=\Delta(b',\sigma) \neq \emptyset$ and
$E_\sigma(b)=true$. This means, by (\ref{eq:E}), $f_\sigma(b)=1$.
Using the assumption that $\mbox{\textbf{LOC}} =
(\mbox{\textbf{LOC}}_{st},\mbox{\textbf{LOC}}_{cf})$ is control
equivalent to $\mbox{\textbf{SUP}}$, in particular
(\ref{eq:con_equiv}), we derive $g_\sigma(i)=1$. Since $b' \in
B_{k,i}$ and $\Delta(b',\sigma) \neq \emptyset$, it follows again
from (\ref{eq:con_equiv}) that $f_\sigma(b')=1$. This implies
$D_\sigma(b')=false$ by (\ref{eq:D}), and therefore $E_\sigma(b)
\wedge D_\sigma(b')=false$. The same argument shows $E_\sigma(b')
\wedge D_\sigma(b)=false$.

Second, if $T(b)=T(b')=0$, then $M(b)=M(b')=0$, and there holds
$T(b)=T(b') \Rightarrow M(b)=M(b')$. Now suppose that
$T(b)=T(b')=1$; by (\ref{eq:T}) $b, b' \models P_m$. Assume on the
contrary that $M(b)=1$ and $M(b')=0$, i.e. $b \models P^f_m$ and $b'
\nvDash P^f_m$ (the other case where $M(b)=0$ and $M(b')=1$ is
similar). By the definition of $I_{m,k}$ in (P1) of the procedure,
above, and $b \in B_{k,i} \cap \{b \in B_S | b \models P^f_m\}$, we
obtain $i \in I_{m,k}$. On the other hand, since
$\mbox{\textbf{LOC}} =
(\mbox{\textbf{LOC}}_{st},\mbox{\textbf{LOC}}_{cf})$ is control
equivalent to $\mbox{\textbf{SUP}}$, it follows from
(\ref{eq:mark_equiv}) and $b' \nvDash P^f_m$ that $i \notin
I_{m,k}$. We have thus derived a contradiction, so $M(b)=M(b')$
after all. \hfill $\square$

\begin{figure}[!t]
  \centering
  \includegraphics[width=0.45\textwidth]{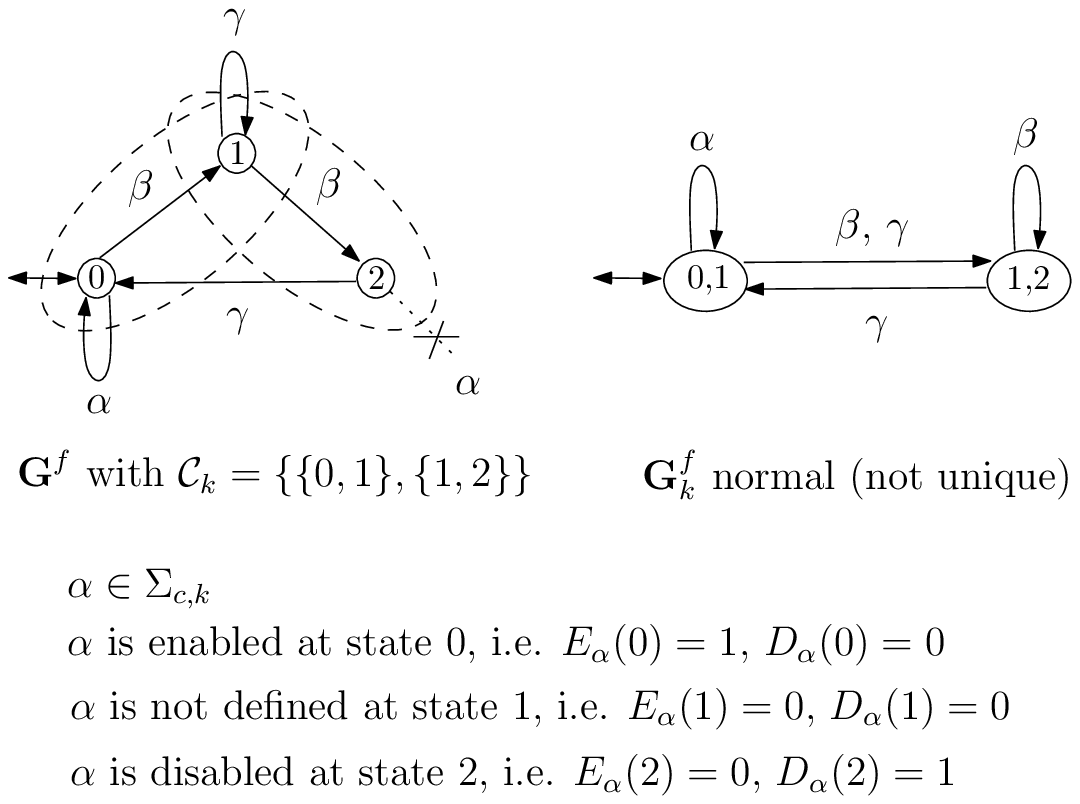}
  \caption{Example: $\textbf{G}^f_k$ constructed from control cover $\mathcal {C}_k = \{\{0,1\},\{1,2\}\}$ is normal with respect to
  $\textbf{G}^f$, control equivalent to $\mbox{\textbf{SUP}}$ with respect to event $\alpha$, and $|\textbf{G}^f_k| < |\textbf{G}^f|$.
  Consider the partitions $\mathcal {C}^1_k = \{\{0,2\},\{1\}\}$, $\mathcal {C}^2_k = \{\{0,1\},\{2\}\}$, and $\mathcal {C}^3_k = \{\{0\},\{1,2\}\}$.
  One verifies that they are not control congruences: for $\mathcal {C}^1_k$ condition (i) of Definition~\ref{defn:concov} fails, and
  for $\mathcal {C}^2_k$, $\mathcal {C}^3_k$ condition (ii) of Definition~\ref{defn:concov} fails. Thus no control congruence can realize $|\textbf{G}^f_k| < |\textbf{G}^f|$.}
  \label{fig:cover}
\end{figure}

That the $\mathcal {C}_k$ are covers in Theorem~\ref{them:sts_1} is
important if the state size of $\textbf{G}^f_k$ is required to be
smaller than that of $\textbf{G}^f$ (as is usually the case in
practice). In particular, if \emph{control cover} is replaced by
\emph{control congruence}, then there may not exist a normal pair
$\mbox{\textbf{LOC}} =
(\mbox{\textbf{LOC}}_{st},\mbox{\textbf{LOC}}_{cf})$ that is control
equivalent to $\mbox{\textbf{SUP}}$ and with $|\textbf{G}^f_k| <
|\textbf{G}^f|$; see Fig.~\ref{fig:cover}.

\subsection{Event Sharing} \label{subsec:share}

So far our STS localization theory has been developed under the
assumption that component agents have pairwise disjoint alphabets
(i.e. Assumption 1).  Now we remove this assumption and discuss the
case where agents may share events.  This also provides an extension
of \cite{CaiWon_TAC10}. Our localization scheme in the event sharing
case is first to synthesize a local state tracker and a local
control function for each controllable event, rather than for each
agent, and then allocate the synthesized local state trackers and
local control functions among the set of agents.

Fix a controllable event $\sigma \in \Sigma_c$. We decompose the
monolithic state tracker $\textbf{G}^f$ into a local state tracker
$\textbf{G}^f_\sigma$ for the event $\sigma$. The decomposition
procedure is the same as before, but with some definitions revised
as follows.  Define $\mathcal {R}_\sigma \subseteq B_S \times B_S$
to be a \emph{control consistency relation} with respect to $\sigma$
by $(\forall b,b' \in B_S)\ (b, b') \in \mathcal {R}_\sigma$ if and
only if
\begin{align*}
&(i)\  E_\sigma(b) \wedge D_\sigma(b') = false = E_\sigma(b') \wedge D_\sigma(b);\\
&(ii)\ T(b)=T(b') \Rightarrow M(b) = M(b').
\end{align*}
Similar to the relation $\mathcal {R}_k$ in
Definition~\ref{defn:consis}, $\mathcal {R}_\sigma$ is reflexive and
symmetric, but need not be transitive, and consequently leads to a
cover on the set $B_S$. Let $I_\sigma$ be some index set, and
$\mathcal {C}_\sigma = \{B_{\sigma,i} \subseteq B_S | i \in
I_\sigma\}$ be a cover on $B_S$. Define $\mathcal {C}_\sigma$ to be
a \emph{control cover} with respect to $\sigma$ by
\begin{align*}
&(i)\ (\forall i \in I_\sigma, \forall b,b' \in B_{\sigma,i})\ (b,b') \in \mathcal {R}_\sigma;  \\
&(ii)\ (\forall i \in I_\sigma, \forall \sigma' \in \Sigma) \Big[
(\exists b \in B_{\sigma,i})\ \Delta^f(b,\sigma') \neq \emptyset
\Rightarrow \\
&\hspace{1cm}(\exists j \in I_\sigma)(\forall b' \in B_{\sigma,i})\
\Delta^f(b',\sigma') \subseteq B_{\sigma,j} \Big].
\end{align*}
Based on a control cover $\mathcal {C}_\sigma$ on $B_S$, we
construct using the procedure (P1)-(P3) in Section~\ref{subsec:proc}
a local state tracker $\textbf{G}^f_\sigma = (I_\sigma,
\Sigma_{l,\sigma}, \delta_\sigma, i_{0,\sigma}, I_{m,\sigma})$ for
the event $\sigma$.  Finally, we define a corresponding local
control function $g_\sigma: I_\sigma \rightarrow \{0,1\}$ by
$g_\sigma(i)=1 \mbox{\ \ if and only if \ \ } (\exists b \in
B_{\sigma,i})\ b \models \Gamma(N_{good}(\sigma), \sigma)$.

Now for each controllable event $\sigma \in \Sigma_c$ we derive a
local state tracker $\textbf{G}^f_\sigma$ and a local control
function $g_\sigma$. Let $\textbf{LOC} = \{(\textbf{G}^f_\sigma,
g_\sigma) \ |\ \sigma \in \Sigma_c\}$ be the set of local
controllers, and $\mbox{\textbf{SUP}} := (\textbf{G}^f,\{f_\sigma |
\sigma \in \Sigma_c\})$ be the optimal and nonblocking monolithic
supervisor. Then we have the following result.
\begin{prop} \label{prop:share}
$\textbf{LOC}$ is control equivalent to $\mbox{\textbf{SUP}}$;
namely, for every $\sigma \in \Sigma_{c}$ and $b \in B_S$, there
exists $i \in I_k$ such that $b \in B_{\sigma,i}$ and
\begin{align*}
& (i)\ \Delta(b,\sigma) \neq \emptyset \ \Rightarrow \ \Big[
f_\sigma(b) = 1
\mbox{ if and only if } g_\sigma(i) = 1 \Big];\\
& (ii)\ b \models P_m^f \mbox{ if and only if } b \models P_m \ \& \
i \in I_{m,k}.
\end{align*}
\end{prop}
\noindent \emph{Proof.} Let $\sigma \in \Sigma_{c}$ and $b \in B_S$.
Then by the definition of control cover $\mathcal {C}_\sigma$ on
$B_S$, there must exist a state $i \in I_\sigma$ of the tracker
$\textbf{G}^f_\sigma$ corresponding to a cell $B_{\sigma,i}$ of the
cover $\mathcal {C}_\sigma$ such that $b \in B_{\sigma,i}$.  The
rest of the proof follows similarly to that of
Theorem~\ref{thm:suploc}. \hfill $\square$

Finally, we allocate the derived local state trackers and local
control functions (with respect to individual controllable events)
among the set of component agents $\textbf{G}_k$, $k \in [1,n]$.
There may be different ways of allocation, allowing case-dependent
choices. For example, if $\textbf{G}_k$ and $\textbf{G}_j$ share a
controllable event $\sigma$, i.e. $\sigma \in \Sigma_{c,k} \cap
\Sigma_{c,j}$, then the local state tracker $\textbf{G}^f_\sigma$
and local control function $g_\sigma$ can be allocated to either
agent or to both.  Allocating to both agents may increase robustness
against faults because even if one fails, the other can continue
operating; on the other hand, allocating to either agent would be
cheaper for implementation. So in practice there is often a tradeoff
between robustness and cost.

Among other others, the following is a convenient allocation, in the
sense that every $\textbf{G}^f_\sigma$ and $g_\sigma$ is implemented
by exactly one agent.
\begin{equation} \label{eq:allocation}
\begin{split}
&\textbf{G}_1: \ \ (\forall \sigma \in \Sigma_{c,1})\ \textbf{G}^f_\sigma, \ g_\sigma\\
&\textbf{G}_2: \ \ (\forall \sigma \in \Sigma_{c,2} \setminus \Sigma_{c,1})\ \textbf{G}^f_\sigma, \ g_\sigma\\
&\hspace{0.15cm} \vdots \\
&\textbf{G}_n: \ \ (\forall \sigma \in \Sigma_{c,n} \setminus
(\Sigma_{c,n-1} \cup \cdots \cup \Sigma_{c,1}) )\
\textbf{G}^f_\sigma, \ g_\sigma
\end{split}
\end{equation}
Choosing this or (obvious) alternative ways of allocation would be
case-dependent.


\section{Symbolic Localization Algorithm} \label{Sec4_LocAlg}

\begin{figure*}[!t]
  \centering
  \includegraphics[width=0.65\textwidth]{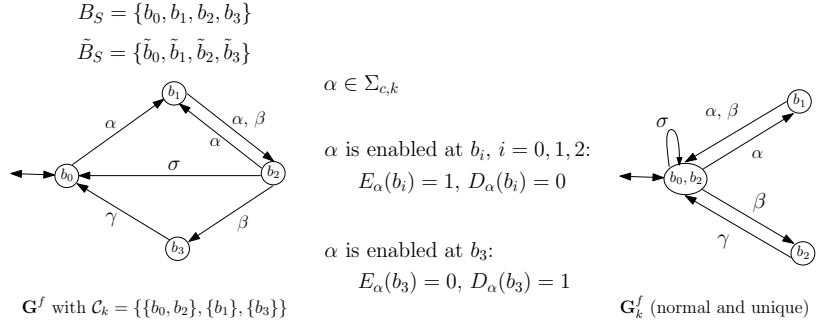}
  \caption{Example: STS localization algorithm}
  \label{fig:sts_ex4}
\end{figure*}

In this section we design an STS localization algorithm for
computing local controllers, which is more efficient than the
counterpart algorithm in \cite{CaiWon_TAC10}.

We have seen in the preceding section that Theorems~\ref{thm:suploc}
and \ref{them:sts_1} together establish the same conclusion as in
the RW framework \cite{CaiWon_TAC10}: namely every set of control
covers generates a solution to the Distributed Control Problem, and
every \emph{normal} solution to the Distributed Control Problem can
be constructed from some set of control covers. In particular, a set
of \emph{state-minimal} local state trackers (possibly non-unique)
can in principle be defined from a set of suitable control covers.
It would thus be desirable to have an efficient algorithm that
computes such a set of covers; however, the minimal state problem is
known to be NP-hard \cite{SuWon:04}. Nevertheless, a polynomial-time
localization algorithm was proposed in \cite{CaiWon_TAC10} which
generates a control congruence (instead of a control cover), and
empirical evidence \cite{CaiWon_IJAMT10} shows that significant
state size reduction can often be achieved. In the following, we
propose a new localization algorithm which is based on STS. The
advantage of using STS is that the efficiency of the new algorithm
is improved compared to the one in \cite{CaiWon_TAC10}, as will be
shown below.

We sketch the idea of the algorithm as follows. Let $B_S$ in
(\ref{eq:sup}) be labeled as $B_S = \{b_0,\ldots,b_{N-1}\}$, and
$\Sigma_{c,k} \subseteq \Sigma_c$ be the controllable events of
agent $\textbf{P}_k$, $k \in [1,n]$.  Our algorithm will generate a
control congruence $\mathcal {C}_k$ on $B_S$ (with respect to
$\Sigma_{c,k}$). This is done symbolically.  First introduce the set
$\tilde{B}_S = \{\tilde{b}_0,\ldots,\tilde{b}_{N-1}\}$, where
$\tilde{b}_i : \mathcal {B}(\textbf{S}T) \rightarrow \{0,1\}$ are
predicates defined by $\tilde{b}_i(b)=1$ if and only if $b=b_i$. Two
elements of $\tilde{B}_S$ may be merged (by ``$\vee$'') if (i) their
corresponding basic trees are control consistent (line 10 in the
pseudocode below, where $\tilde{\mathcal {R}}_k : Pwr(B_S)
\rightarrow \{0,1\}$ is defined by $B_1 \models \tilde{\mathcal
{R}}_k$ if and only if $(\forall b,b' \in B_1) (b,b') \in \mathcal
{R}_k$); and (ii) all corresponding downstream basic trees reachable
from $b, b'$ by identical strings are also control consistent (line
12, where $\tilde{\Delta}:Pred(\textbf{S}T) \times \Sigma
\rightarrow Pred(\textbf{S}T)$ is the predicate counterpart of
$\Delta$ in (\ref{eq:sts_tree})). We note that since
$\tilde{\Delta}$ can handle one-step transitions of a predicate
corresponding to a subset of basic trees, in each call of the
CHECK\_MERGE function we may also check control consistency by
applying $\tilde{\mathcal {R}}_k$ to this subset; this is more
efficient than the algorithm in \cite{CaiWon_TAC10} which in each
call of the CHECK\_MERGE function checks control consistency only
for a pair of flat states (corresponding to basic trees). Finally,
after checking all the elements in $\tilde{B}_S$, the algorithm at
line 8 generates a control congruence $\mathcal {C}_k$ each cell of
which consists of the basic trees $b_i$ whose corresponding
predicates $\tilde{b}_i$ are merged together in $\tilde{B}_S$.

\begin{thm} \label{thm:localg}
The STS localization algorithm terminates, has (worst-case) time
complexity $O(N^3)$, and the generated $\mathcal {C}_k$ is a control
congruence on $B_S$.
\end{thm}

Before proving Theorem~\ref{thm:localg}, we remark that the STS
localization algorithm realizes the same functionality as the one in
\cite{CaiWon_TAC10}, and moreover improves the time complexity from
$O(N^4)$ in \cite{CaiWon_TAC10} to $O(N^3)$. This is achieved by the
fact that the (global) transition function of STS can handle subsets
of basic trees simultaneously, which makes checking the control
consistency relation in each call of the CHECK\_MERGE function more
efficient.

The following is the pseudocode of the algorithm.  Notation:
``$\setminus$'' denotes set subtraction; $x \prec y$ means $x
\preceq y$ and $x \neq y$.

\begin{algorithmic}[1]
\Procedure{main()}{}

\For{$i := 0$ to $N-2$}

    \For{$j := i+1$ to $N-1$}

        \State $B = \tilde{b}_i \vee \tilde{b}_j$;

        \State $W = \emptyset$;

        \If{Check$\_$Merge($B, W, i, \tilde{B}_S$) = $true$}
        \State $\tilde{B}_S = (\tilde{B}_S \cup W) \setminus \{\tilde{b} \in \tilde{B}_S \ |\ (\exists w \in W) \tilde{b} \prec w\}$; \EndIf

    \EndFor

\EndFor

\State \textbf{return} $\mathcal {C}_k = \{ \cup_i b_{i} \ |\ \vee_i
\tilde{b}_i \in \tilde{B}_S \}$;

\EndProcedure


\Function{Check$\_$Merge}{$B, W, i, \tilde{B}_S$}

\If{$\{b \in \mathcal {B}(\textbf{S}T) \ |\ b \models B\} \nvDash
\tilde{\mathcal {R}}_k$} \textbf{return} $false$; \EndIf

\State $W = (W \cup B) \setminus \{w \in W \ |\ w \prec B\}$;

\For{each $\sigma \in \Sigma$ with $\tilde{\Delta}(B, \sigma) \wedge
S \neq false$}

            \If{$(\tilde{\Delta}(B, \sigma) \wedge S) \preceq w$ for some $w \in W \cup \tilde{B}_S$} \textbf{continue}; \EndIf

            \If{$(\tilde{\Delta}(B, \sigma) \wedge S) \wedge \tilde{b}_r \neq false$ for some $r < i$}
            \textbf{return} $false$; \EndIf

            \State $B = (\tilde{\Delta}(B, \sigma) \wedge
            S) \vee (\bigvee \{w | w \in W \ \&\ w \wedge (\tilde{\Delta}(B, \sigma) \wedge S) \neq false \})$;

            \If{Check$\_$Merge($B, W, i, \tilde{B}_S$) = $false$}
            \textbf{return} $false$; \EndIf

\EndFor

\State \textbf{return} $true$;

\EndFunction
\end{algorithmic}

\noindent \emph{Proof of Theorem~\ref{thm:localg}.} Since both $B$
at line 4 and $\tilde{\Delta}(B, \sigma) \wedge S$ at line 15 are
the join ``$\vee$'' of the predicates in $\tilde{B}_S$, so is each
element of $W$ which is updated only at line 11.  Thus, the size of
$\tilde{B}_S$, which is updated only at line 7, is non-increasing.
Because the initial size $N$ is finite, the algorithm must
terminate. In the worst case, there can be $N(N-1)/2$ calls (by
lines 2, 3) made to the function CHECK\_MERGE, which can then make
$N$ calls (by lines 12, 13) to itself.  So the worst-case time
complexity is $N^2(N-1)/2 = O(N^3)$.

It is left to show that $\mathcal {C}_k$ generated at line 8 is a
control congruence. First, the control consistency of every pair of
basic trees in the same cell of $\mathcal {C}_k$ is guaranteed by
the check at line 10; so $\mathcal {C}_k$ is a control cover.
Second, the set subtraction ``$\setminus$'' when updating $W$ at
line 11 and $\tilde{B}_S$ at line 7 ensures that the cells of
$\mathcal {C}_k$ are pairwise disjoint; thus $\mathcal {C}_k$ is a
partition on $B_S$. Therefore, we conclude that $\mathcal {C}_k$ is
a control congruence. \hfill $\square$

\begin{figure*}[!t]
  \centering
  \includegraphics[width=0.9\textwidth]{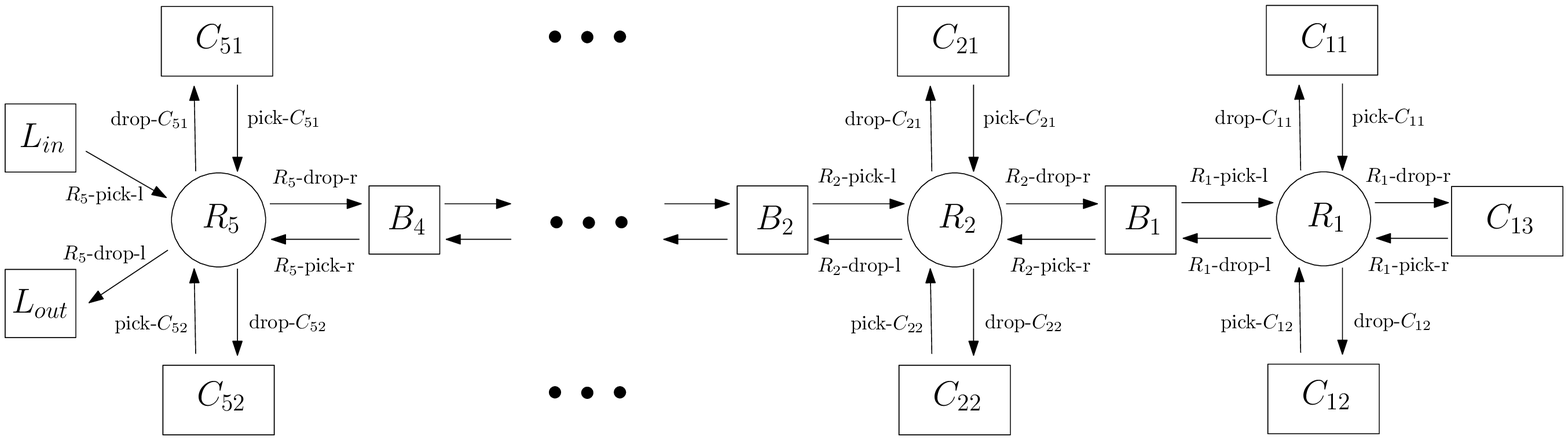}
  \caption{Cluster Tool: an integrated semiconductor manufacturing system used
for wafer processing.}
  \label{fig:clutol}
\end{figure*}

\begin{exmp}\label{sts_ex4}

We provide an example, displayed in Fig.~\ref{fig:sts_ex4}, to
illustrate the STS localization algorithm. Initially, $\tilde{B}_S
=\{\tilde{b}_0,\tilde{b}_1, \tilde{b}_2, \tilde{b}_3\}$. The ranges
of indices $i$ and $j$ at lines 2 and 3 are $i \in [0,2]$ and
$j\in[i+1,3]$, respectively.

(1) $(\tilde{b}_0,\tilde{b}_1)$ cannot be merged. First,
$B=\tilde{b}_0 \vee \tilde{b}_1$ and the test at line 10 is passed
since $\{b_0,b_1\}\models \tilde{\mathcal {R}}_k$; so $W=\tilde{b}_0
\vee \tilde{b}_1$. Second, $B$ is updated at line 15 to
$B=\tilde{b}_0 \vee \tilde{b}_1 \vee \tilde{b}_2$ and the test at
line 10 is still passed since $\{b_0,b_1,b_2\}\models
\tilde{\mathcal {R}}_k$; so $W=\tilde{b}_0 \vee \tilde{b}_1 \vee
\tilde{b}_2$. Third, $B$ is updated at line 15 to $B=\tilde{b}_0
\vee \tilde{b}_1 \vee \tilde{b}_2 \vee \tilde{b}_3$ but now the test
at line 10 fails since $\{b_0,b_1,b_2,b_3\}\nvDash \tilde{\mathcal
{R}}_k$ (indeed, $(b_i,b_3) \notin \mathcal {R}_k$, $i=0,1,2$). Note
that when $B=\tilde{b}_0 \vee \tilde{b}_1 \vee \tilde{b}_2$ the
global transition function $\tilde{\Delta}(B, \sigma)$ at lines
12-15 handles the local transitions at basic trees $b_1,b_2,b_3$
simultaneously:
\[\tilde{\Delta}(\tilde{b}_0 \vee \tilde{b}_1 \vee \tilde{b}_2,
\alpha) = \tilde{b}_0 \vee \tilde{b}_1 \vee \tilde{b}_2 \ \ \ \&\ \
\ \tilde{\Delta}(\tilde{b}_0 \vee \tilde{b}_1 \vee \tilde{b}_2,
\beta) = \tilde{b}_2 \vee \tilde{b}_3.\] This operation is more
efficient than the localization algorithm in \cite{CaiWon_TAC10};
there, only a pair of basic trees of $\{b_1,b_2,b_3\}$ and the
associated transitions can be processed at a single step.

(2) $(\tilde{b}_0,\tilde{b}_2)$ can be merged. First, $B=\tilde{b}_0
\vee \tilde{b}_2$ and the test at line 10 is passed since
$\{b_0,b_2\}\models \tilde{\mathcal {R}}_k$; so $W=\tilde{b}_0 \vee
\tilde{b}_2$. Second, $B$ is updated at line 15 to $B=\tilde{b}_1$
and the test at line 10 is trivially passed; so $W=\{\tilde{b}_0
\vee \tilde{b}_2, \tilde{b}_1\}$. Now one verifies that the
condition at line 13 is satisfied for both transitions $\alpha$ and
$\beta$ defined at $b_1$, so the ``for''-loop from line 12 to line
16 is finished without calling the CHECK\_MERGE function. Hence
$true$ is returned at line 6 and $\tilde{B}_S$ is updated at line 7
to $\tilde{B}_S=\{\tilde{b}_0 \vee \tilde{b}_2, \tilde{b}_1,
\tilde{b}_3\}$.

(3) $(\tilde{b}_0,\tilde{b}_3)$ cannot be merged because $(b_0,b_3)
\notin \mathcal {R}_k$ and the test at line 10 fails.

(4) $(\tilde{b}_1,\tilde{b}_2)$ cannot be merged. First,
$B=\tilde{b}_1 \vee \tilde{b}_2$ and the test at line 10 is passed
since $\{b_1,b_2\}\models \tilde{\mathcal {R}}_k$; so $W=\tilde{b}_1
\vee \tilde{b}_2$. Second, $B$ is updated at line 15 to
$B=\tilde{b}_1 \vee \tilde{b}_2 \vee \tilde{b}_3$ but the test at
line 10 fails since $\{b_1,b_2,b_3\}\nvDash \tilde{\mathcal {R}}_k$.

(5) $(\tilde{b}_1,\tilde{b}_3)$ cannot be merged because $(b_1,b_3)
\notin \mathcal {R}_k$ and the test at line 10 fails.

(6) $(\tilde{b}_2,\tilde{b}_3)$ cannot be merged because $(b_2,b_3)
\notin \mathcal {R}_k$ and the test at line 10 fails.

Finally, $\tilde{B}_S=\{\tilde{b}_0 \vee \tilde{b}_2, \tilde{b}_1,
\tilde{b}_3\}$ and line 8 generates a control congruence $\mathcal
{C}_k = \{\{b_0, b_2\}, \{b_1\}, \{b_3\}\}$. The normal local state
tracker (unique in this case) constructed from $\mathcal {C}_k$ is
displayed in Fig.~\ref{fig:sts_ex4}.
\end{exmp}


\section{Case Study Cluster Tool} \label{Sec5_Appli}

In this section, we demonstrate STS supervisor localization on
Cluster Tool, an integrated semiconductor manufacturing system used
for wafer processing (e.g. \cite{YiDiZhMe:07}). Starting with a
decentralized approach (see \cite{CGWW:12} for a recent development
in STS), we apply localization to establish a purely distributed
control architecture for Cluster Tool. A second purpose of this case
study is to compare our results with those reported in
\cite{SuSchu:10,SuSchu:12} for the same (structured) system;  by
imposing additional specifications we derive straightforward
coordination logic, which provides global nonblocking control.

\begin{figure}[!t]
  \centering
  \includegraphics[width=0.45\textwidth]{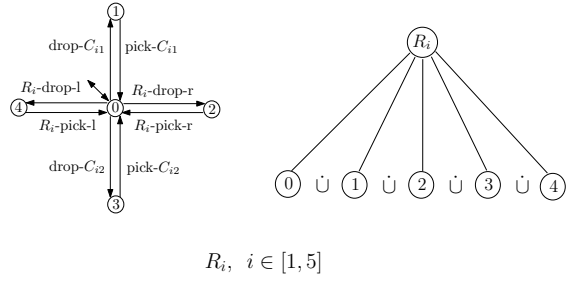}
  \caption{Plant components}
  \label{fig:Ri}
\end{figure}

As displayed in Fig.~\ref{fig:clutol}, Cluster Tool consists of (i)
two loading docks ($L_{in}, L_{out}$) for wafers entering and
leaving the system, (ii) eleven vacuum chambers ($C_{11}, C_{12},
\ldots, C_{52}$) where wafers are processed, (iii) four buffers
($B_1, \ldots, B_4$) where wafers are temporarily stored, and (iv)
five robots ($R_1, \ldots, R_5$) which transport wafers in the
system along the following production sequence:
\begin{align*}
L_{in} \rightarrow C_{51} \rightarrow B_4 \rightarrow \cdots
\rightarrow B_2 \rightarrow C_{21} \rightarrow B_1 \rightarrow
C_{11} &\downarrow \\ &C_{13}\\ L_{out} \leftarrow C_{52} \leftarrow
B_4 \leftarrow \cdots \leftarrow B_2 \leftarrow C_{21} \leftarrow
B_1 \leftarrow C_{12} &\downarrow.
\end{align*}
The five robots are the plant components; their automaton dynamics
and state trees are displayed in Fig.~\ref{fig:Ri}. Each robot $R_i$
has $8$ events, all assumed controllable.  Note that the robots have
pairwise disjoint alphabets; thus Assumption 1 at the beginning of
Section~\ref{Sec3_ProFor} is satisfied.

\begin{figure*}[!t]
  \centering
  \includegraphics[width=0.75\textwidth]{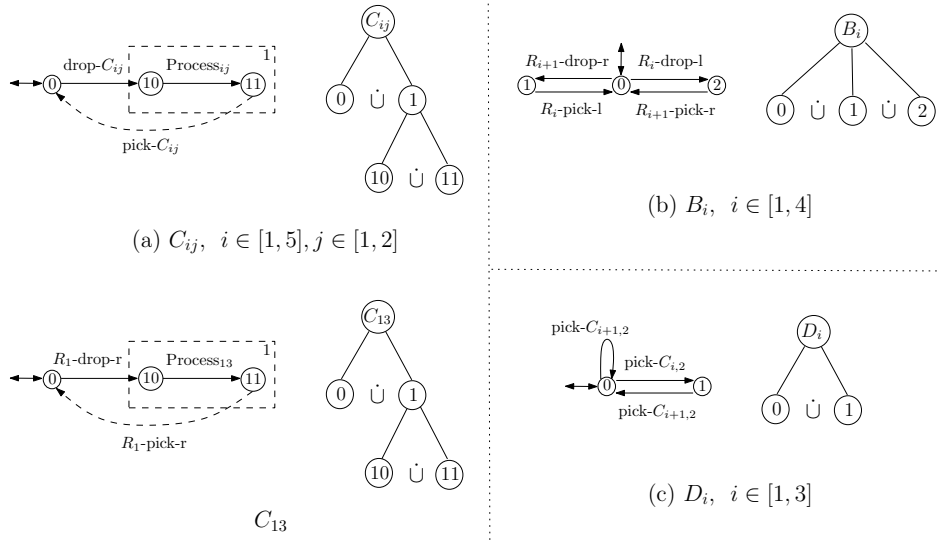}
  \caption{Control specifications}
  \label{fig:Bi}
\end{figure*}

Next, we describe control specifications for Cluster Tool. (1)
Fig.~\ref{fig:Bi}~(a): at each chamber $C_{ij}$ a wafer is first
dropped in by robot $R_i$, then processed, and finally picked up by
$R_i$. Thus a chamber behaves essentially like a one-slot buffer;
our first control specification is to protect each $C_{ij}$ against
overflow and underflow. Note also that the event Process$_{ij}$
(designated uncontrollable) can be viewed as an internal transition
of chamber $C_{ij}$; so for its corresponding two states ``$10$''
and ``$11$'', we introduce a hierarchy in $C_{ij}$'s state tree
model. (2) Fig.~\ref{fig:Bi}~(b): each buffer $B_i$ has capacity
one, and may be incremented by $R_i$ from the right (resp. $R_{i+1}$
from the left) and then decremented by $R_{i+1}$ from the left
(resp. $R_i$ from the right). Our second control specification is to
protect all buffers against overflow and underflow. Thus far we have
described specifications related to physical units -- chambers and
buffers; the final requirement, denoted by $D_i$ ($i\in[1,3]$), is
purely logical, and coordinates the operations between neighboring
robots. (3) Fig.~\ref{fig:Bi}~(c): once robot $R_i$, $i\in[1,3]$,
picks up a wafer from chamber $C_{i,2}$, it may not do so again
until robot $R_{i+1}$ empties chamber $C_{i+1,2}$. The rationale for
imposing this specification is as follows (refer to
Fig.~\ref{fig:clutol}): once a wafer is picked up by $R_i$,
$i\in[1,3]$, it needs to be transported through $B_i \rightarrow
R_{i+1} \rightarrow C_{i+1,2} \rightarrow R_{i+1} \rightarrow
B_{i+1}$; here buffers $B_i, B_{i+1}$ and robot $R_{i+1}$ can be
viewed as shared resources, and if chamber $C_{i+1,2}$ is full, then
the above wafer transportation may cause blocking. Hence a
reasonable (but possibly restrictive) requirement to avoid system
deadlock is to guarantee an empty slot in $C_{i+1,2}$ before $R_i$
initiates the wafer transportation.
Note that we do not impose the same specification between $R_4$ and
$R_5$, because $R_5$ can drop wafers out of the system without
capacity constraint.

\begin{figure}[!t]
  \centering
  \includegraphics[width=0.45\textwidth]{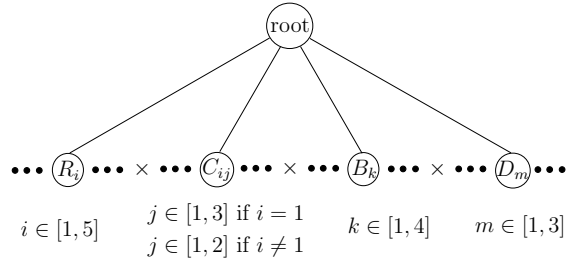}
  \caption{State tree model $\textbf{S}T_{CT}$ of Cluster Tool}
  \label{fig:CT_ST}
\end{figure}

Putting together plant components (Fig.~\ref{fig:Ri}) and control
specifications (Fig.~\ref{fig:Bi}), we obtain the state tree model
$\textbf{S}T_{CT}$ of Cluster Tool, displayed in
Fig.~\ref{fig:CT_ST}. The system is large-scale -- the
(uncontrolled) total state size is approximately $3.6 \times
10^{11}$. Moveover, apart from satisfying all imposed control
specifications, the system will require nontrivial coordination to
prevent deadlocks caused by conflicts in using multiple shared
resources (robots and buffers). Consequently, the overall optimal
and nonblocking control of Cluster Tool is a challenging design
exercise.

Directly applying the monolithic supervisor synthesis of STS
\cite{MaWon:05,MaWon:06TAC} results in an optimal and nonblocking
supervisor $S$ (predicate as in (\ref{eq:sup})); the corresponding
global state tracker has $3227412$ basic trees, and the associated
control functions of certain events have a large number of BDD
nodes:
\begin{align*}
&f_{\mbox{pick-}C_{12}}: 205 \mbox{ nodes},\ \
f_{\mbox{pick-}C_{21}}: 284 \mbox{ nodes},\\
&f_{\mbox{pick-}C_{22}}: 319 \mbox{ nodes},\ \
f_{\mbox{pick-}C_{31}}: 686 \mbox{ nodes},\\
&f_{\mbox{pick-}C_{32}}: 571 \mbox{ nodes},\ \
f_{\mbox{pick-}C_{41}}: 1561 \mbox{ nodes},\\
&f_{\mbox{pick-}C_{42}}: 777 \mbox{ nodes},\ \
f_{\mbox{pick-}C_{51}}: 5668 \mbox{ nodes}.
\end{align*}
Because of the large sizes, it is difficult to grasp the control
logic, and to implement the state tracker and control functions in
practice.

\begin{figure*}[!t]
  \centering
  \includegraphics[width=1.0\textwidth]{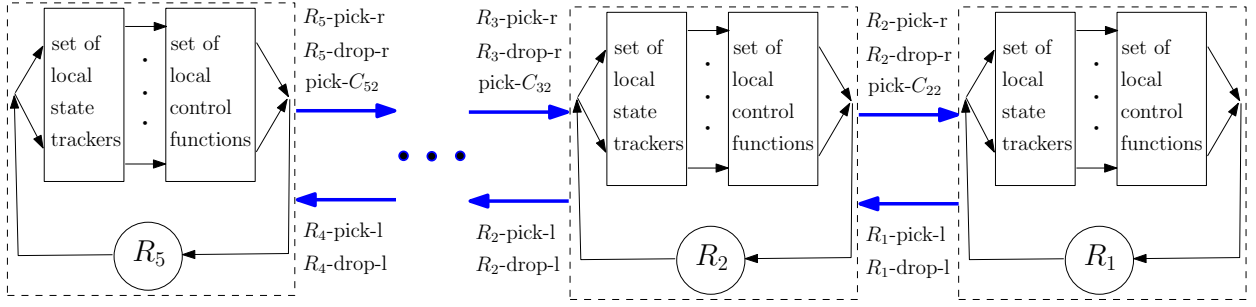}
  \caption{Distributed control architecture for Cluster Tool: each robot is supervised by its own set of local state trackers
  and local control functions, as well as interacting (e.g. through event communication) with its immediate left and right neighbors.}
  \label{fig:CT_st_dm}
\end{figure*}

Our synthesis goal is to derive, by applying STS supervisor
localization, a set of local state trackers and local control
functions for each of the five robots such that (1) the
corresponding control logic is transparent, and moreover (2) the
collective local control action is identical to the monolithic
optimal and nonblocking control action of $S$. Specifically, we will
derive the distributed control architecture displayed in
Fig.~\ref{fig:CT_st_dm}; as will be shown, the
interconnection/communication among robots involves only nearest
neighbors.

\begin{rem}
The setup of our Cluster Tool is borrowed from
\cite{SuSchu:10,SuSchu:12}, except for the following. (1) Our system
has one more robot (and the corresponding buffer and chambers), so
the total state size is of order $10^3$ larger than the system size
in \cite{SuSchu:10,SuSchu:12}. (2) The control specifications $D_i$,
$i\in[1,3]$ (Fig.~\ref{fig:Bi}(c)), were not imposed in
\cite{SuSchu:10,SuSchu:12}. The specifications $D_i$ make the
system's behavior more restrictive; as we shall see, however, with
$D_i$ imposed we derive straightforward control and coordination
logics which achieve global optimal and nonblocking supervision,
despite the larger state size of our case. In addition, as explained
above, these specifications $D_i$ are themselves reasonable
requirements to prevent system deadlock.
\end{rem}

\textbf{Decentralized control and supervisor localization.} To
facilitate applying STS supervisor localization on the large system
at hand, we use a decentralized approach.  Since each control
specification (Fig.~\ref{fig:Bi}) relates (in the sense of sharing
events) to no more than two plant components (robots), a
corresponding optimal and nonblocking decentralized supervisor may
be synthesized as in (\ref{eq:sup}). Then the developed localization
algorithm is applied to decompose each decentralized supervisor. In
this way both the supervisor synthesis and the localization may be
based merely on the relevant robot(s), thereby making the
computations more efficient. Concretely, we proceed as follows.

\begin{figure*}[!t]
  \centering
  \includegraphics[width=0.9\textwidth]{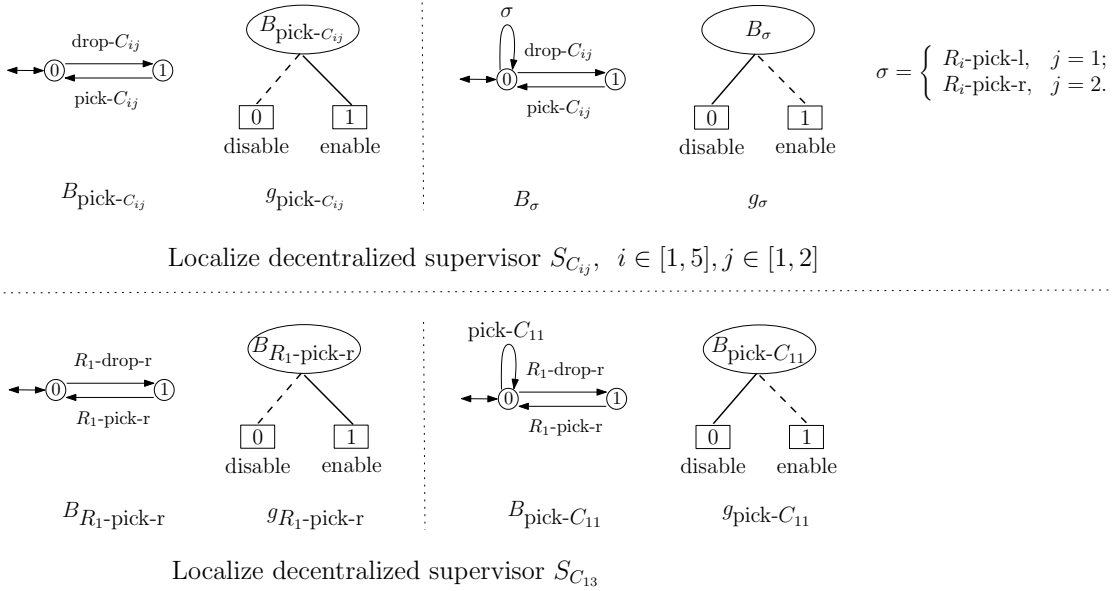}
  \caption{Local state trackers and local control functions obtained by localizing decentralized supervisors $S_{C_{ij}}$.
  Here each state tracker (e.g. $B_{\mbox{pick-}C_{ij}}$) has $2$ states, which are encoded by one BDD node in the corresponding
  control function (e.g. $g_{\mbox{pick-}C_{ij}}$) taking binary
  values either $0$ (displayed by dashed line) or $1$ (solid line).}
  \label{fig:SUP_Ci}
\end{figure*}

(1) Chamber specifications $C_{ij}$ (Fig.~\ref{fig:Bi}(a)). Each
$C_{ij}$, $i\in[1,5]$, shares events only with robot $R_i$. Thus we
treat $R_i$ as plant, $C_{ij}$ as specification, and compute an
optimal and nonblocking decentralized supervisor $S_{C_{ij}}$
(predicate as in (\ref{eq:sup})); the corresponding state tracker
has $8$ states. Then we apply our algorithm to localize each
$S_{C_{ij}}$, and obtain a set of local state trackers and local
control functions for the relevant controllable events, as displayed
in Fig.~\ref{fig:SUP_Ci}. For each $S_{C_{ij}}$ there are two events
requiring control action. We explain control logic for the case
where $i\in[1,5]$ and $j=1$; the other cases are similar. One such
event is pick-$C_{ij}$, which must be disabled ($R_i$ may not pick
up a wafer from chamber $C_{ij}$) if $C_{ij}$ is empty; this is to
protect chamber $C_{ij}$ against underflow. The other event
requiring control action is $R_i$-pick-l, which must be disabled
($R_i$ may not pick up a wafer from left) if chamber $C_{ij}$ is
full. This rule prevents a deadlock situation: if $R_i$ took a wafer
and $C_{ij}$ were full, then $R_i$ could neither drop the wafer to
$C_{ij}$ nor pick up a wafer from $C_{ij}$. The rule at the same
time prevents chamber $C_{ij}$ from overflow.  Note that controlling
just event drop-$C_{ij}$ suffices to prevent overflow, but cannot
prevent deadlock.

\begin{figure*}[!t]
  \centering
  \includegraphics[width=0.9\textwidth]{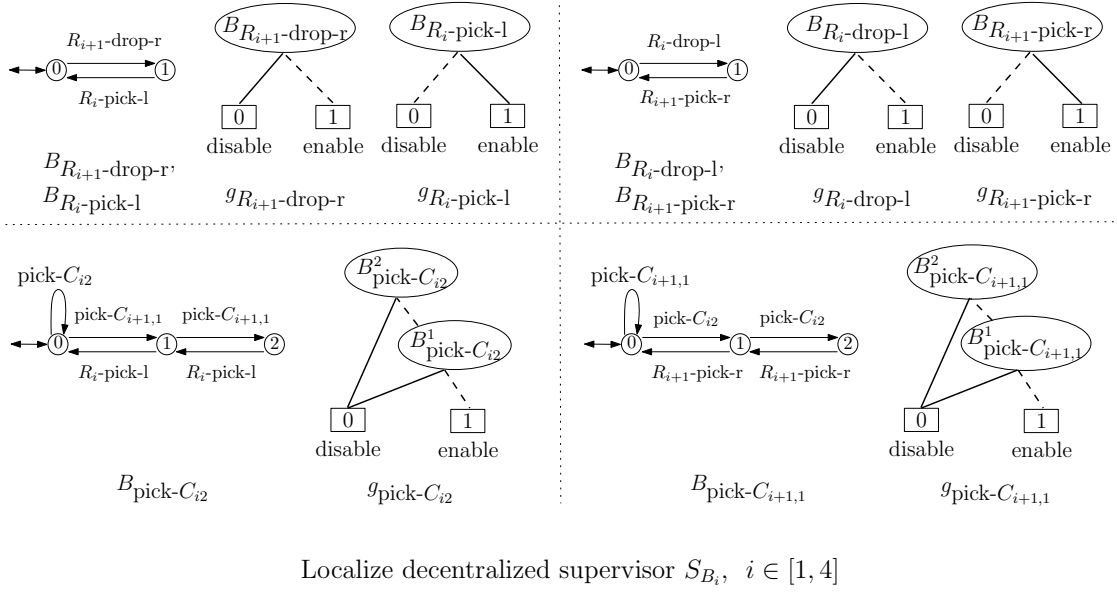}
  \caption{Local state trackers and local control functions obtained by localizing decentralized supervisors $S_{B_{i}}$.
  Here the state tracker $B_{\mbox{pick-}C_{i2}}$ (resp. $B_{\mbox{pick-}C_{i+1,1}}$) has $3$ states, which are encoded by two BDD nodes in the corresponding
  control function $g_{\mbox{pick-}C_{i2}}$ (resp. $g_{\mbox{pick-}C_{i+1,1}}$). For example, state $2$ of $B_{\mbox{pick-}C_{i2}}$ is encoded as $(B^2_{\mbox{pick-}C_{i2}}, B^1_{\mbox{pick-}C_{i2}})=(1,0)$.}
  \label{fig:SUP_Bi}
\end{figure*}

(2) Buffer specifications $B_{i}$ (Fig.~\ref{fig:Bi}(b)). Each
$B_{i}$, $i\in[1,4]$, shares events with two robots, $R_i$ and
$R_{i+1}$. Treating $R_i$ and $R_{i+1}$ as plant, $B_{i}$ as
specification, we compute a decentralized supervisor $S_{B_{i}}$
(predicate as in (\ref{eq:sup})); the corresponding state tracker
has $55$ states. Localizing each $S_{B_{i}}$, we obtain a set of
local state trackers and associated control functions for the
relevant controllable events, as displayed in Fig.~\ref{fig:SUP_Bi}.
For each $S_{B_{i}}$ there are six events requiring control action.
Events $R_{i}$-drop-l and $R_{i+1}$-drop-r must be disabled ($R_{i}$
or $R_{i+1}$ may not drop a wafer into buffer $B_i$) when $B_i$ is
full -- this is to prevent buffer overflow. On the other hand,
events $R_{i}$-pick-l and $R_{i+1}$-pick-r must be disabled ($R_{i}$
or $R_{i+1}$ may not pick up a wafer from buffer $B_i$) when $B_i$
is empty -- this is to prevent buffer underflow.  In addition to
preventing buffer overflow and underflow, event pick-$C_{i2}$ must
be disabled ($R_i$ may not pick up a wafer from chamber $C_{i2}$)
unless there is no wafer on the path $R_{i}$-$B_i$-$R_{i+1}$. This
logic is to prevent the deadlock situation where both $R_i$ and
$R_{i+1}$ pick up a wafer to transport through $B_i$, but neither
can do so because the buffer has capacity of only one. For the same
reason, event pick-$C_{i+1,1}$ must be disabled ($R_{i+1}$ may not
pick up a wafer from chamber $C_{i+1,1}$) unless there is no wafer
on the path $R_{i}$-$B_i$-$R_{i+1}$.

\begin{figure}[!t]
  \centering
  \includegraphics[width=0.45\textwidth]{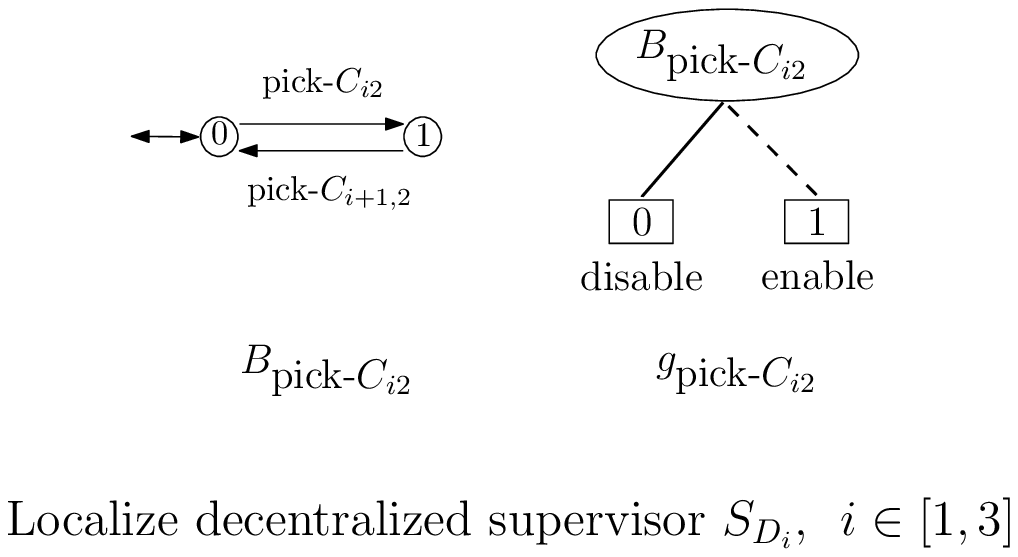}
  \caption{Local state trackers and local control functions obtained by localizing decentralized supervisors $S_{D_{i}}$.}
  \label{fig:SUP_Di}
\end{figure}

(3) Specifications $D_{i}$ (Fig.~\ref{fig:Bi}(c)). Like buffer
specifications, each $D_{i}$, $i\in[1,3]$, shares events with robots
$R_i$ and $R_{i+1}$.  Treating $R_i$ and $R_{i+1}$ as plant, $D_{i}$
as specification, we first synthesize a decentralized supervisor
$S_{D_{i}}$ (the corresponding state tracker has $50$ states), and
then apply localization to compute a set of local state trackers and
associated control functions for the relevant controllable events,
as displayed in Fig.~\ref{fig:SUP_Di}. For each $S_{D_{i}}$ only the
event pick-$C_{i2}$ requires control action: it must be disabled
($R_i$ may not pick up a wafer from chamber $C_{i2}$) if the
neighboring chamber $C_{i+1,2}$ is full. This logic is to prevent
blocking while wafers are transported from right to left in the
system, as we explained above when the specifications were imposed.

\textbf{Coordination logic and STS verification.} We have obtained a
set of decentralized supervisors $S_{C_{ij}}$ ($i\in[1,5]$,
$j\in[1,3]$ if $i=1$ and $j\in[1,2]$ otherwise), $S_{B_k}$
($k\in[1,4]$), and $S_{D_m}$ ($m\in[1,3]$). Viewing these
decentralized supervisors as predicates defined on the set $\mathcal
{B}(\textbf{S}T_{CT})$ of basic trees of $\textbf{S}T_{CT}$, we
define their joint behavior $S_{joint}$ by
\begin{align*}
S_{joint} := \left( \bigwedge_{i,j} S_{C_{ij}} \right) \wedge \left(
\bigwedge_{k} S_{B_{k}} \right) \wedge \left( \bigwedge_{m}
S_{D_{m}} \right).
\end{align*}
Unfortunately $S_{joint}$ is not the same as the monolithic
supervisor $S$. Indeed, there exists \emph{conflict} among the
decentralized supervisors, so that there are non-coreachable basic
trees satisfying the joint behavior $S_{joint}$. This is verified as
follows\footnote{Algorithm~3 in \cite{CGWW:12} can also be used for
the verification.}: let $S_{joint}$ be the plant (so the
decentralized supervisors' state trackers are the plant components),
let the predicate $true$ be the specification (i.e. no additional
control requirement is imposed), and compute the corresponding
optimal and nonblocking supervisor. This supervisor turns out to be
the same as the monolithic $S$, but there are four events --
pick-$C_{i1}$, $i\in[2,5]$ -- requiring disablement action; their
control functions have BDD node sizes $14$, $78$, $471$, and $2851$,
respectively. Since no new control constraint was imposed, the above
required disablement action controllably removes the blocking basic
trees of $S_{joint}$ and thereby reproduces the nonblocking
monolithic $S$.

\begin{figure*}[!t]
  \centering
  \includegraphics[width=0.85\textwidth]{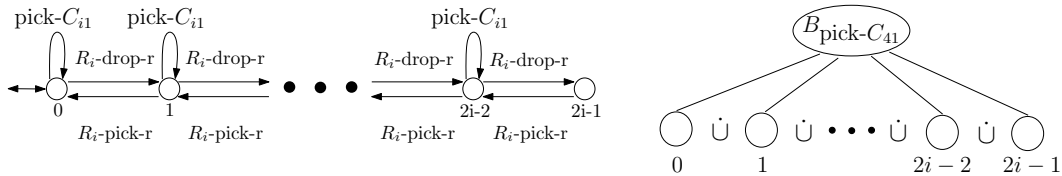}
  \caption{Coordinators CO$_{\mbox{pick-}C_{i1}}$, $i\in[2,5]$, each having $2i$ states.}
  \label{fig:Coord}
\end{figure*}

\begin{table*}[!t]
\renewcommand{\arraystretch}{1.3}
\caption{State/BDD node size comparison between distributed and
monolithic approaches.} \label{tab:number} \centering
\begin{tabular}{|c|cccc|c|}\hline
\multirow{3}{*}{Events} & \multicolumn{4}{|c|}{State sizes of local state trackers localized from} & {BDD node numbers of} \\
 & \multicolumn{4}{|c|}{decentralized supervisors and of coordinators} & {control functions of} \\
 & $S_{C_{ij}}$ in Fig.~\ref{fig:SUP_Ci} & $S_{B_{i}}$ in Fig.~\ref{fig:SUP_Bi} & $S_{D_{i}}$ in Fig.~\ref{fig:SUP_Di} & $\mbox{CO}_{\mbox{pick-}C_{i1}}$ in Fig.~\ref{fig:Coord} & monolithic supervisor $S$ \\
\hline
pick-$C_{12}$ & 2 & 3 & 2 & --- & 205 \\
pick-$C_{21}$ & 2 & 3 & --- & 4 & 284 \\
pick-$C_{22}$ & 2 & 3 & 2 & --- & 319 \\
pick-$C_{31}$ & 2 & 3 & --- & 6 & 686 \\
pick-$C_{32}$ & 2 & 3 & 2 & --- & 571 \\
pick-$C_{41}$ & 2 & 3 & --- & 8 & 1561 \\
pick-$C_{43}$ & 2 & 3 & --- & --- & 777 \\
pick-$C_{51}$ & 2 & 3 & --- & 10 & 5668 \\
\hline
\end{tabular}
\end{table*}

One could use as \emph{coordinators} the above computed control
functions of events pick-$C_{i1}$, $i\in[2,5]$. Because of the large
BDD node sizes of certain events (especially pick-$C_{41}$ and
pick-$C_{51}$), however, the control logic is still difficult to
grasp. Instead we propose, based on analyzing the structure of
Cluster Tool and the wafer transportation route
(Fig.~\ref{fig:clutol}), the coordinators CO$_{\mbox{pick-}C_{i1}}$,
$i\in[2,5]$, as displayed in Fig.~\ref{fig:Coord}. We explain the
coordination logic. Observe in Fig.~\ref{fig:clutol} that once a
wafer is picked up from chamber $C_{i1}$ (i.e. pick-$C_{i1}$
occurs), it will be transported by robot $R_i$ to the right, and so
all the way to $R_1$ and then back to the left to $R_i$ -- a looping
route. For example, when $R_3$ takes a wafer from $C_{31}$, the loop
is as follows:
\begin{align*}
C_{31} \underrightarrow{\ \ \ R_3\ \ \ } B_2  \underrightarrow{\ \ \
R_2\ \ \ } C_{21}
 \underrightarrow{\ \ \ R_2\ \ \ } B_1  \underrightarrow{\ \ \ R_1\ \ \ }
C_{11} &\downarrow\ R_1 \\
&C_{13}\\
C_{32} \overleftarrow{\ \ \ R_3\ \ \ } B_2 \overleftarrow{\ \ \ R_2\
\ \ } C_{21} \overleftarrow{\ \ \ R_2\ \ \ } B_1 \overleftarrow{\ \
\ R_1\ \ \ } C_{12} &\downarrow\ R_1.
\end{align*}
Since the loop has limited capacity to hold wafers, control is
needed at the entrance and exit of the loop to prevent `choking' the
loop with too many wafers.  The logic of the coordinators in
Fig.~\ref{fig:Coord} specifies that event pick-$C_{i1}$ must be
disabled if the number of wafers input exceeds wafers output by
$2i-1$. Note that the loop capacity $2i-1$ is exactly the number of
chambers in the loop; this is because robots and buffers are shared
resources, and if all the chambers are full, inputting one more
wafer to the loop will clearly cause deadlock. We remark that the
proposed coordination rule requires global knowledge; for example,
to disable event pick-$C_{51}$, the coordinator
CO$_{\mbox{pick-}C_{51}}$ needs to know \emph{a priori} the capacity
of the whole loop on the right. Upon knowing the loop capacity,
however, each coordinator may be implemented locally because it
suffices just to count the numbers of wafers input and output to the
corresponding loop.

We now verify that the four proposed coordinators
CO$_{\mbox{pick-}C_{i1}}$, $i\in[2,5]$ (Fig.~\ref{fig:Coord}),
indeed resolve all the conflicts among the decentralized
supervisors. Let the four coordinators and the decentralized
supervisors' state trackers be the plant components, the predicate
$true$ be the specification, and compute the corresponding optimal
and nonblocking supervisor. This supervisor turns out to be the same
as the monolithic supervisor $S$, and now no event requires further
control action. This computation shows that the proposed
coordinators and the decentralized supervisors together provide the
same global optimal and nonblocking control action as the monolithic
supervisor $S$ did.

On the other hand, by Theorem~\ref{thm:suploc} each pair comprising
a local state tracker and a local control function in
Figs.~\ref{fig:SUP_Ci}-\ref{fig:SUP_Di} is control equivalent to the
corresponding decentralized supervisor. Therefore, the set of
controllers and coordinators in
Figs.~\ref{fig:SUP_Ci}-\ref{fig:Coord} is control equivalent to the
monolithic supervisor $S$.  Finally grouping them with respect to
the individual robots $R_i$, $i\in[1,5]$, we derive the distributed
control architecture displayed earlier in Fig.~\ref{fig:CT_st_dm}
where each robot interacts only with its nearest neighbor(s).

\begin{rem}
In the work of \cite{SuSchu:10,SuSchu:12} on Cluster Tool, the
primary focus was on reducing computational complexity in achieving
global optimal and nonblocking control. There the authors proposed
an efficient ``distributed supervisor'' synthesis, based on
abstraction and coordination techniques, which solves the Cluster
Tool problem by involving state sizes of order only $10^2$ in the
computations.  It is not clear, however, what the resulting control
and coordination rules are.  Engineers, on the other hand, expect
comprehensible rules for easy implementation and safe management,
especially when the plant itself has an intelligible structure; in
this case, the system components are connected in a loop.

In our results, by contrast, every control/coordination rule is
transparent, as displayed in Figs.~\ref{fig:SUP_Ci}-\ref{fig:Coord}.
The control rules are derived by applying our developed STS
supervisor localization; the coordination logic is designed by
analyzing the loop structure of the system. As a comparison to the
monolithic result computed above, we see from Table~\ref{tab:number}
that substantial size reduction is achieved by supervisor
localization and coordination design. Finally, for verification that
the derived control and coordination action is globally optimal and
nonblocking, we rely on the computational power of STS and BDD.
\end{rem}

\section{Conclusions} \label{Sec6_Concl}

To solve a distributed control problem of discrete-event systems, we
have developed the top-down supervisor localization approach in the
STS framework.  The approach establishes a purely distributed
control architecture, in which every active agent is endowed with
its own local state trackers and local control functions, while
being coordinated with its fellows through event communication in
such a way that the collective local control action is identical to
the global optimal and nonblocking action. Such a control scheme
facilitates distributed and embedded implementation of control
strategies into individual agents.  Compared to the language-based
RW counterpart \cite{CaiWon_TAC10}, we have designed a more
efficient symbolic localization algorithm by exploiting BDD
computation. Furthermore, we have demonstrated our localization
approach in detail on a complex semiconductor manufacturing system,
Cluster Tool.


\bibliographystyle{IEEEtran}
\bibliography{dist_cont,decen_sup,fundam,DistributedControl}

\end{document}